\documentclass{article}

\usepackage{PRIMEarxiv}
\usepackage{float}
\usepackage{braket}
\usepackage{amsmath}
\usepackage[utf8]{inputenc} % allow utf-8 input
\usepackage[T1]{fontenc}    % use 8-bit T1 fonts
\usepackage{hyperref}       % hyperlinks
\usepackage{url}            % simple URL typesetting
\usepackage{booktabs}       % professional-quality tables
\usepackage{amsfonts}       % blackboard math symbols
\usepackage{nicefrac}       % compact symbols for 1/2, etc.
\usepackage{microtype}      % microtypography
\usepackage{lipsum}
\usepackage{fancyhdr}       % header
\usepackage{graphicx}       % graphics
\graphicspath{{media/}}     % organize your images and other figures under media/ folder

\title{Effects of One-particle Reduced Density Matrix Optimization in Variational Quantum Eigensolvers
\thanks{\textit{\underline{Citation}}: 
\textbf{This is a preprint. The final version will be available upon publication.}} 
}

\author{
  Amanda Marques de Lima \\
  Departamento de Química Fundamental \\
  Universidade Federal de Pernambuco \\
  Recife-PE, Brazil\\
  \texttt{amanda.aml002@gmail.com} 
  \And
  Erico Souza Teixeira \\
  Centro de Exelência em Computação Quântica, \\
  Venturus, \\
  Campinas-SP, Brazil \\
  \texttt{erico.teixeira@venturus.org.br} 
  \And
  Eivson Darlivam Rodrigues de Aguiar Silva \\
  Departamento de Química Fundamental \\
  Universidade Federal de Pernambuco \\
  Recife-PE, Brazil\\
  \texttt{eivsondras@gmail.com} 
  \And
  Ricardo Luiz Longo \\
  Departamento de Química Fundamental \\
  Universidade Federal de Pernambuco \\
  Recife-PE, Brazil\\
  \texttt{ricardo.longo@ufpe.br} 
}

\begin{document}
\maketitle

\begin{abstract}
The variational quantum eigensolver (VQE) is a promising method for simulating molecular systems on near-term quantum computers. This approach employs energy estimation; however, other relevant molecular properties can be extracted from the one-particle reduced density matrix (1-RDM) generated by VQE. The accuracy of these properties strongly depends on the reliability and convergence of the 1-RDM, which is not guaranteed by energy-only optimization. Thus, we investigate the effect of optimizing the 1-RDM within VQE to improve the accuracy of both the energy and molecular properties. A two-step algorithm was implemented that optimizes the energy and 1-RDM by incorporating a penalty term in the cost function to enforce the convergence of the 1-RDM. The first step focuses on energy minimization, while in the second step, a weighted penalty is added to the cost function to promote simultaneous improvement of the energy and 1-RDM. This approach was tested and validated for the k-UpCCGSD and GateFabric \textit{ansätzes} with active spaces (4,4) and (2,2), respectively. k-UpCCGSD produces energies close to CISD, so optimizing 1-RDM has little effect on the energy but significantly improves electronic properties such as electron density, dipole moments, and atomic charges. GateFabric initially shows higher energy deviations from CISD, but optimizing 1-RDM substantially improves both the energy accuracy and the quality of 1-RDM. These results demonstrate that simultaneous optimization of energy and 1-RDM is an effective strategy to improve the accuracy of energies and molecular properties in variational quantum algorithms.
\end{abstract}

% keywords can be removed

\section{Introduction}

Quantum mechanics is fundamental for the accurate description of molecular properties such as electron energy, charge density, and dissociation profiles \cite{nakatsuji2015free, jensen2017introduction}. These properties can be obtained from the solution of the Schrödinger equation. However, the exact solution of this equation for systems with many electrons is computationally challenging, which motivated the development of approximate methods \cite{jensen2017introduction}.

Despite their success, these methods often come with a computational cost that grows exponentially with the size and complexity of the system \cite{chaganti2024execution}. This limitation is evident in systems with strong electron correlation, such as those containing transition metals. For example, the accurate description of the FeMo cofactor can require petabytes of memory and impractical execution times \cite{reiher2017elucidating, lee2023evaluating, marti2024spin, feldmann2024complete}. Despite significant progress in classical computing, further advancements are increasingly limited by fundamental physical constraints, including saturation of transistor scaling, memory bandwidth, storage capacity bottlenecks, and limited scalability of parallel algorithms due to Amdahl’s Law and interconnect latency \cite{nofer2023quantum}.  

These limitations are particularly evident in the simulation of quantum systems, where, as Feynman argued \cite{feynman1982simulating}, "nature itself is governed by quantum mechanics, suggesting that more natural and efficient simulations should be carried out on devices that also obey quantum laws". In this scenario, quantum computers have emerged as a promising alternative to overcome such limitations \cite{yang2023survey}. Although these improvements are still theoretical and depend on advances in \textit{hardware} and algorithms, these computers offer a compact representation of complex systems. For example, while the classical simulation of the caffeine molecule would require approximately 1048 bits, a quantum computer could represent the same system with approximately 160 quantum bits, or qubits \cite{vasques2024machine}.

In this context, one of the most promising applications of quantum computing is the simulation of chemical systems. This has the potential to transform areas such as medicine, materials science, catalysis, and nanotechnology \cite{babbush2023quantum, weidman2024quantum}. To this end, several quantum algorithms have been proposed, with Quantum Phase Estimation \cite{kitaev1995quantum}, or QPE, being one of the best known. However, applying QPE to chemically relevant systems typically requires fault-tolerant quantum computers with millions of logical qubits and deep quantum circuits, which makes it impractical for current devices of the NISQ (Noisy Intermediate-Scale Quantum) era \cite{de2022survey, preskill2018quantum}.

Alternatively, the Variational Quantum Eigensolver (VQE) stands out as a hybrid algorithm that combines measurements on quantum computers with optimizations on classical computers \cite{peruzzo2014variational, cerezo2021variational}. VQE allows the estimation of the ground state energy of molecules with greater practical feasibility than QPE, having been successfully tested in systems such as protocatechuic acid \cite{de2024evaluating}, benzene \cite{sennane2023benzene}, and diazene and hydrogen chains with up to 12 atoms \cite{google2020hartree}.

Despite these advances, VQE still faces challenges in accuracy due to the dependence of the results on the quantum circuit (sequence of logic gates applied to each qubit) and the search space accessed during optimization \cite{fedorov2022vqe}. So, improvements to this algorithm are being sought \cite{weidman2024quantum}.

An important direction for improving VQE involves moving beyond energy as the sole metric of performance. Although the primary goal of VQE is to minimize the total energy, this does not ensure that the one-particle reduced density matrix (1-RDM) also converged satisfactorily. This could limit the appropriate description of properties more sensitive to fluctuations of the 1-RDM such as forces, dipole moments, and Mulliken populations \cite{yuan2022quantification, gibney2022density, szabo1996modern}. Therefore, enhancing the estimation of the 1-RDM within VQE is essential for obtaining more reliable predictions of molecular properties, especially in systems with strong electron correlation.

Current developments of VQE have been in two directions: using properties derived from the 1-RDM as validation metrics, or exploring alternative strategies for optimizing the 1-RDM. In the first category, Skogh et al. \cite{skogh2024electron} proposed validating VQE by comparing experimentally accessible properties derived from 1-RDM, such as electron density topologies and atomic partial charges, rather than comparing to 1-RDM directly. Similarly, Le et al. \cite{le2023correlated} employed properties, such as dipole moments, also derived from 1-RDM, as benchmarks for assessing the quality of the VQE results.

In contrast to approaches that use properties derived from 1-RDM as validation metrics, a second category of research focuses on incorporating density matrices into iterative refinement schemes, more closely resembling classical methods. For instance, Tilly et al. \cite{tilly2021reduced} integrated VQE into a complete active space self-consistent field (CASSCF) framework, where the active space wavefunction and molecular orbitals are iteratively optimized using the two-particle reduced density matrix (2-RDM) obtained from VQE. This 2-RDM is used to update, classically, the orbital rotation matrix, enabling variational mixing of core and virtual orbitals. The molecular Hamiltonian is then reconstructed in the new orbital basis, and VQE is rerun until convergence is achieved. Using this approach, the authors obtained accurate estimates of dipole moments and Fermi liquid parameters. Similarly, Lew et al. \cite{lew2025efficient} employed VQE to generate a wavefunction and extract the 1-RDM, whose diagonalization yielded natural orbitals. These orbitals were then used to transform the integrals, enabling the calculation of energy through a natural orbital functional (NOF).

Traditional electronic structure methods, such as those implemented in PySCF \cite{sun2018pyscf} and Gaussian \cite{frish2009gaussian} codes, often employ convergence criteria based on the root-mean-square deviation (RMSD) of the 1-RDM between iterations as well as the maximum deviation. Inspired by this classical approach, we incorporate a similar RMSD-based term into the cost function of VQE. By weighting the RMSD between consecutive 1-RDMs together with the energy, our method promotes the simultaneous convergence of the energy and electron density, aiming at increasing the accuracy of molecular properties. Compared to recent iterative orbital optimization schemes, our approach offers a simpler alternative, as it preserves the original VQE structure, avoiding the need for iterative Hamiltonian reconstructions or full diagonalization of the 1-RDM. Furthermore, it provides a more accessible means to improve the description of molecular properties in cases where energy minimization alone does not guarantee wavefunction convergence.

Thus, this work modified the traditional VQE flow to incorporate 1-RDM optimization, together with energy. For this purpose, a code capable of extracting these properties from 1-RDM optimized by VQE was also developed. The results were compared to those obtained with non-optimized 1-RDM.

Section II introduces the fundamental concepts of VQE and 1-RDM, detailing how molecular properties can be derived from this matrix. It also presents the procedure for obtaining and optimizing the 1-RDM within the VQE framework, together with the techniques employed in this work, both on classical computers and quantum simulators. Section III presents and discusses the results obtained for the CH$_5^+$ molecule, highlighting the benefits of the proposed approach. Section IV summarizes the main findings and outlines potential directions for future research.

\section{Methods}

This section describes the computational methods employed to investigate the impact of optimizing the 1-RDM within the VQE framework. The standard VQE algorithm, including its mathematical formulation, circuit design, and \textit{ansätze} used for quantum state preparation is initially outlined. Then, the formalism of the 1-RDM and its connection to chemically relevant properties such as electron density, dipole moment, and Mulliken charges is presented. A modified VQE approach that simultaneously minimizes energy and refines the 1-RDM is introduced, which enables improved prediction of molecular properties. The systems of interest are presented with the simulation parameters, and the active space configurations used in both standard and enhanced VQE implementations.

\subsection{Variational Quantum Eigensolver}

In the VQE algorithm, the cost function is given by the expectation value of the Hamiltonian of the system, $\hat{H}$, evaluated over a parameterized trial wavefunction, in accordance with the Schrödinger equation:
\begin{equation}
\hat{H} | {\Psi_0(\theta)} \rangle = E_0 |\Psi_0 (\theta)\rangle,
\end{equation}
where $E_0$ is the ground-state energy, and $|\Psi_0(\theta)\rangle$ is a variational parameterized quantum state that approximates the eigenvector corresponding to the lowest eigenvalue $E_0$ \cite{cerezo2021variational}. The goal of VQE is to adjust the variational parameters $\theta$ to minimize the expectation value of $\hat{H}$, providing an accurate estimate of the ground-state energy. This procedure is based on the variational principle, ensuring that the estimated energy will always be greater than or equal to the exact energy of the ground state \cite{tilly2022variational}.

\begin{figure}[ht!]
\centering
\includegraphics[width=1.0\textwidth]{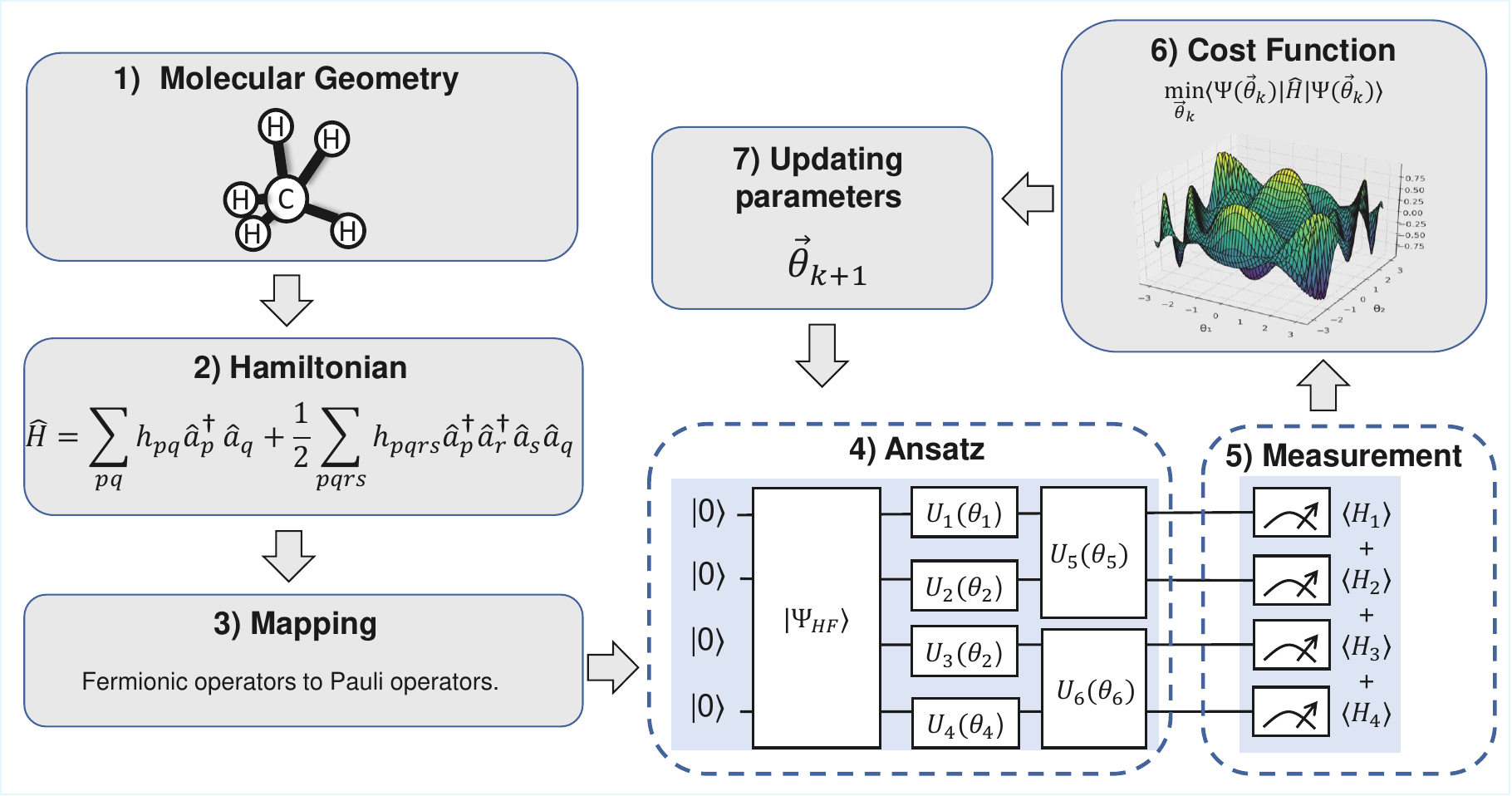}
\\
\caption{Representation of the flow of the VQE algorithm. The steps in gray refer to the processes executed on the classical computer.}
\label{fig:vqe-structure}
\end{figure}

The steps of the VQE algorithm are represented in Figure \ref{fig:vqe-structure}. Initially, for a given (optimized) molecular geometry, the electronic Hamiltonian of the molecule is obtained in the second quantized form. This Hamiltonian is expressed in the fermion basis and, therefore, cannot be directly implemented in qubit-based quantum devices. To make it compatible, it is necessary to transform it to the Pauli operator basis \cite{cao2019quantum, fedorov2022vqe}.

This transformation can be performed through different mappings, such as the Jordan–Wigner (JW), Bravyi-Kitaev, and Parity schemes. Among them, the JW mapping is widely used because it offers a direct correspondence between spin-orbitals and qubits. In this case, occupied and unoccupied spin-orbitals are mapped to the qubit states $|1\rangle$ and $|0\rangle$, respectively \cite{tranter2018comparison}. As a result, the molecular fermionic Hamiltonian is mapped onto a quibt Hamiltonian which is expressed in terms of Pauli strings \cite{fedorov2022vqe}:

\begin{equation}
\hat{H} = \sum_{j} \alpha_{j} P_{j},
\end{equation}

where $\alpha_j$ are real coefficients and $P_j$ represent products of Pauli operators acting on the qubits.

Once the Hamiltonian in the Pauli basis is obtained, the next step consists of choosing a parameterized \textit{ansatz}, in addition to defining the initial state and the variational parameters $\theta$. Typically, a Hartree-Fock (HF) wavefunction is used as the initial state \cite{fedorov2022vqe}. When the initial parameters $\theta$ are zero, the prepared state exactly coincides with the HF state. However, if the parameters are initialized randomly, the initial state will deviate from the HF configuration \cite{fedorov2022vqe}.

The quantum computer (or simulator) is then used to estimate the expectation value of $\hat{H}$ with respect to the parametrized state $|\Psi(\theta)\rangle$. This process involves measurements on the qubits to calculate the average of each term $P_j$ that makes up the Hamiltonian \cite{fedorov2022vqe}. The parameters of the \textit{ansatz} are iteratively updated by a classical optimizer, aiming at minimizing the estimated energy \cite{fedorov2022vqe, mcardle2020quantum}. At each iteration, the quantum device evaluates the energy using the updated parameters, continuing this process until a convergence criterion for the energy is achieved \cite{fedorov2022vqe, mcardle2020quantum}. Then, the optimized parameters yield an estimate of the ground-state energy of the system and the wavefunction can be employed in obtaining molecular properties.

In molecular energy calculations, chemistry-inspired \textit{ansätze} have been widely adopted from the experiences with electronic structure methods. Notable examples include the Unitary Coupled Cluster truncated to single and double excitations (UCCSD) \cite{romero2018strategies} and its extension, the k-Unitary Pair Coupled Cluster Generalized Singles and Doubles (k-UpCCGSD) \cite{lee2018generalized}. In contrast, hardware-efficient \textit{ansätze}, such as GateFabric \cite{anselmetti2021local}, are designed to minimize circuit depth and adapt to the connectivity constraints of near-term quantum devices, while preserving key physical symmetries, including particle number and total spin.

The k-UpCCGSD is an extension of the UCCSD, created to reduce the number of gates in the circuit by introducing two main modifications \cite{lee2018generalized}. First, the single and double excitations are generalized, that is, any orbitals can be excited, without distinguishing between occupied and virtual ones. Second, the double excitations follow the pair coupled-cluster (pCC) structure, involving only the joint movement of electron pairs between orbitals, which reduces the number of terms in the circuit \cite{lee2018generalized}. 

The ansatz is applied repeatedly in a layered fashion, where the number of layers is denoted by the prefactor k. Each layer contains a distinct set of excitation operators, and stacking multiple layers enhances the expressiveness of the ansatz. As a result, the cost of preparing a k-UpCCGSD state scales linearly with the system size, multiplied by the prefactor k, providing a tunable trade-off between circuit depth and accuracy. Increasing k systematically improves the quality of the approximation, enabling better convergence to the true ground state \cite{fedorov2022vqe}.

In turn, GateFabric uses Givens rotations to perform single and double excitations, in addition to adding fermionic SWAP gates. Furthermore, this circuit performs double excitations only between neighboring qubits \cite{anselmetti2021local}. In addition to its hardware-efficient design, GateFabric preserves important physical symmetries, such as particle number and total spin, making it well-suited for simulating fermionic systems \cite{arrazola2022universal, anselmetti2021local}.

\subsection{Reduced Density Matrix and Molecular Properties}

In the second quantization formalism, the expectation value of any operator $\hat{\Omega}$ can be expressed as \cite{helgaker2013book}:

\begin{equation}
\langle \Psi | \hat{\Omega} | \Psi \rangle = \sum_{pq} {D}_{pq} \Omega_{pq} + \frac{1}{2} \sum_{pqrs} {d}_{pqrs} \Omega_{pqrs} + \Omega_0,
\end{equation}

where $\Omega_{pq}$ and $\Omega_{pqrs}$ represent the one-electron and two-electron integrals, respectively, and $\Omega_0$ is a constant term. The quantities $D_{pq}$ and $d_{pqrs}$ correspond to the one- and two-particle reduced density matrices (1-RDM and 2-RDM), which encapsulate information about the electronic structure of the system.

The 1-RDM provides insight into the distribution and occupation of molecular orbitals, and is defined as \cite{townsend2019post, helgaker2013book}:

\begin{equation}
D_{pq} = \langle \Psi | a_p^\dagger a_q | \Psi \rangle,
\label{eq-rdm1}
\end{equation}

where $a_p^\dagger$ and $a_q$ are the electron creation and annihilation operators, respectively. This matrix describes the probability of an electron transitioning from the state $q$ to the state $p$, reflecting both the occupancy and the coherence between the orbitals.

This 1-RDM is in the Molecular Orbital (MO) basis, but can be converted to the atomic orbital basis, since the relationship between the MO and the Atomic Orbital (AO) is given by:

\begin{equation} \phi_p(r) = \sum_{\nu} C_{\nu p} \varphi_\nu(r), \label{eq:func_bases_atomicas} \end{equation}

where $C_{\nu p}$ are the molecular coefficients that express the molecular orbitals $\phi_p(r)$ in terms of the atomic basis functions $\varphi_\nu(r)$. Thus, the 1-RDM was converted from the MO basis to the AO basis through the transformation:

\begin{equation}
    \gamma_{\mu \nu} = \sum_{pq} C_{\mu p} D_{pq} C_{\nu q}^*,
    \label{eq-rdm1_base_atomica}
\end{equation}

where $C$ is the coefficient matrix that maps MO to AO, obtained from the HF method using PySCF.

Several molecular properties can be obtained from 1-RDM, such as the electron density \cite{dominguez2024electron}, the molecular electrostatic potential \cite{suresh2022molecular}, the dipole moment \cite{szabo1996modern}, as well as the Mulliken charges and populations \cite{szabo1996modern}. The methods for obtaining these properties from 1-RDM will be discussed in more detail below.

\subsubsection{Electronic Density}

The electron density describes the distribution of electrons in a molecule in space \cite{domingo2016molecular}. It represents the probability of finding an electron at a given point and is a central concept in quantum chemistry. Electron density is not only used to visualize the shape of molecules and chemical bonds but also serves as the foundation for interpreting chemical structure, reactivity, and bonding patterns \cite{gadre2021electrostatic}. High-density regions typically occur near atomic nuclei and between bonded atoms, reflecting areas of strong electron localization. 

The electron density can be obtained from the 1-RDM using the following expression \cite{dominguez2024electron, szabo1996modern}:

\begin{equation} \rho(r) = \sum_{pq} D_{pq} \phi_p^*(r) \phi_q(r), \label{eq-electron_density} \end{equation}
where $D_{pq}$ are the elements of the 1-RDM, and $\phi_p(\mathbf{r})$ and $\phi_q(\mathbf{r})$ represent the basis functions associated with the molecular orbitals.

However, to calculate the electron density, the 1-RDM must first be expressed in the AO basis. This transformation is necessary because the basis functions used by PySCF, the framework employed to evaluate these properties, are defined in the AO representation, while the 1-RDMs obtained from both VQE and HF are originally represented in the Molecular Orbital (MO) basis. 

Thus, With the 1-RDM expressed in the atomic basis, the electron density in the atomic basis could be calculated according to:
\begin{equation}
    \rho(r) = \sum_{\mu \nu} \gamma_{\mu \nu} \varphi_{\mu}^* (r) \varphi_\nu (r),
\end{equation}

where $\varphi_{\mu}^* (r)$ and $ \varphi_\nu (r)$ are the atomic basis functions.

A robust theoretical approach for analyzing chemical structure and reactivity based on electron density is the Quantum Theory of Atoms in Molecules (QTAIM) \cite{cortes2023introduction}. This theory performs a topological analysis of $\rho(\mathbf{r})$, allowing the identification of regions of high electron concentration around nuclei or between pairs of neighboring nuclei \cite{cortes2023introduction}.

The topological characterization is based on the identification of critical points (CPs), defined as the locations where the gradient of the electron density vanishes \cite{skogh2024electron}. These points are classified according to the local curvature of the density, which is described by the Hessian matrix of $\rho(\mathbf{r})$, as shown in Equation~\ref{eq-hessian-matrix} \cite{cortes2023introduction}. Diagonalizing this matrix yields eigenvalues that determine whether a critical point corresponds to a maximum, minimum, or saddle point.

\begin{equation}
\nabla \nabla^T \rho(\mathbf{r}) =
\begin{bmatrix}
\frac{\partial^2 \rho}{\partial x^2} & \frac{\partial^2 \rho}{\partial x \partial y} & \frac{\partial^2 \rho}{\partial x \partial z} \\
\frac{\partial^2 \rho}{\partial y \partial x} & \frac{\partial^2 \rho}{\partial y^2} & \frac{\partial^2 \rho}{\partial y \partial z} \\
\frac{\partial^2 \rho}{\partial z \partial x} & \frac{\partial^2 \rho}{\partial z \partial y} & \frac{\partial^2 \rho}{\partial z^2}
\end{bmatrix}
\label{eq-hessian-matrix}
\end{equation}

Among the main types of critical points, the bond critical points (BCPs) are particularly important. These are located between two nuclei and are indicative of the presence of a chemical bond \cite{cortes2023introduction}. They are characterized by two negative and one positive eigenvalue of the Hessian, reflecting two directions of electron density decrease and one of increase. Nuclear critical points (NCPs), found at the positions of atomic nuclei, correspond to local maxima in electron density and are typically associated with a positive Laplacian, indicating high electron concentration around the nucleus \cite{cortes2023introduction}. Other types of critical points, such as ring and cage critical points, also exist, but will not be discussed in this work.

\subsubsection{Electrostatic Potential}

The electrostatic potential represents the electric field generated by the distribution of nuclear and electronic charges in a molecule \cite{politzer2021molecular}. It describes how a test positive charge would experience force at any point in space around the molecule. This potential plays a central role in understanding chemical reactivity, non-covalent interactions, and molecular recognition, as regions of high or low potential influence how molecules interact with each other \cite{suresh2022molecular}. 

The molecular electrostatic potential generated at a point $\mathbf{r}$ in the vicinity of a molecular system is given by \cite{suresh2022molecular}:

\begin{equation}
V(\mathbf{r}) = \sum_{A} ^N \frac{Z_A}{|\mathbf{r} - \mathbf{R}_A|} - \int \frac{\rho(\mathbf{r}')}{|\mathbf{r} - \mathbf{r}'|} d\mathbf{r}',
\label{eq-potencial_elet}
\end{equation}

where $Z_A$ is the charge of the nucleus $A$, \( \rho(\mathbf{r}') \) is the electron density, obtained from the 1-RDM, while \( |\mathbf{r} - \mathbf{r}'| \) is the distance between the point \( \mathbf{r} \), whose potential is being calculated, and the point \( \mathbf{r}' \), where the charge \( \rho(\mathbf{r}') \) is located.

\subsubsection{Dipole Moment}

The dipole moment is a vector quantity that measures the separation of positive and negative charges within a molecule. It provides a quantitative description of a molecule's polarity, i.e., how asymmetric the charge distribution is. Molecules with large dipole moments tend to interact more strongly with electric fields and polar solvents. 

The molecular dipole moment can be calculated from the 1-RDM using the following expression \cite{szabo1996modern}:

\begin{equation}
    \mu_x = - \sum_{\mu} \sum_{\nu}\gamma_{\mu \nu}  (\nu |\mathbf{x}|\mu) + \sum_A Z_A X_A,
    \label{eq-dip_moment}
\end{equation}

where $\mu_x$ is the $\mathbf{x}$ component of the total dipole moment vector of the molecule, $\gamma_{\mu \nu}$ are the elements of the 1-RDM, $(\nu | \mathbf{x} | \mu)$ represents the integral of the position operator in the $x$ direction, $Z_A$ is the charge of the nucleus $A$ and $X_A$ its $x$ coordinate. The $y$ and $z$ components follow the same analogical form, allowing the construction of the complete dipole vector.

\subsubsection{Mulliken Population Analysis}

Population analyses aim to divide the electron density of a molecule into specific atomic contributions \cite{davidson2022viewpoint}. Partial atomic charges serve as an important rationalization of organic and inorganic reactivity, and are the basis for understanding non-covalent interactions, through the permanent or induced polarization of the electron density \cite{davidson2022viewpoint}.

Mulliken analysis allows estimating the electron populations and partial charges on atoms from the 1-RDM and the overlap matrix, defined as $S_{\nu\mu} = \langle \chi_\nu | \chi_\mu \rangle$, where $\chi_\nu$ and $\chi_\mu$ are basis functions centered on different atoms \cite{davidson2022viewpoint, szabo1996modern}.

The electron population associated with an atom $A$ is given by \cite{szabo1996modern}:

\begin{equation}
P_A = \sum_{\mu \in A} \sum_{\nu} \gamma_{\mu \nu} S_{\nu \mu},
\label{eq-mulliken_pop}
\end{equation}

where \( \mu \in A \) indicates that the \( \mu\) orbital is centered on the \( A \) atom. The Mulliken charge on the \( A \) atom is then obtained by \cite{szabo1996modern}:

\begin{equation}
    q_A = Z_A - P_A,
    \label{eq-charge_mulliken}
\end{equation}

where \( Z_A \) is the nuclear charge of the atom \( A \) and \( P_A \) is the electronic population associated with that atom.

\subsection{1-RDM with Active Space}

Modeling the electronic structure with quantum computing is challenging because the qubit requirements scale linearly with the size of the basis set \cite{fedorov2022vqe}. A common solution is to use small basis sets and the active space approximation, which selects a subset of Molecular Orbitals (MOs) \cite{fedorov2022vqe}, as realized in the CASSCF \cite{levine2020casscf}. A well-constructed, effective Hamiltonian of a small active space can capture the effects of the entire orbital space \cite{metcalf2020resource}. This space typically includes the most relevant MOs, so that MOs that do not belong to the active space remain frozen \cite{sayfutyarova2017automated}.

\begin{figure}[ht!]
    \centering
    \includegraphics[width=0.5\textwidth]{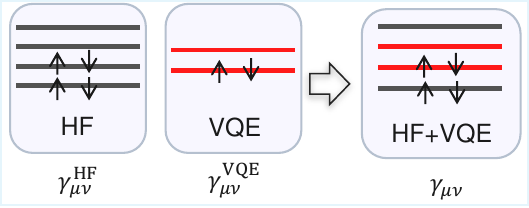}
    \\
    \caption{Construction of 1-RDM using active space. The active and frozen orbitals are generated by VQE and HF, respectively.}
    \label{fig:dpq}
\end{figure}

When an active space is employed in a VQE calculation, the resulting 1-RDM contains elements only for the active orbitals. As shown in the Figure \ref{fig:dpq}, to reconstruct the full molecular 1-RDM, including both active and frozen orbitals, it is necessary to merge the VQE output ($\gamma^{\text{VQE}}_{\mu \nu}$) with information from a HF calculation ($\gamma^{\text{HF}}_{\mu \nu}$). This is achieved by replacing the elements of the HF 1-RDM corresponding to the active orbitals with those computed via VQE, while retaining the HF values for the frozen orbitals (Equation \ref{eq:rdm1-merge}). In this hybrid approach, the electron correlation within the active space is captured by VQE, whereas the frozen orbitals remain described by the HF wavefunction.

\begin{equation}
\gamma_{\mu \nu} = \begin{cases} \gamma^{\text{VQE}}_{\mu \nu}, & \text{if } \mu, \nu \in \text{active space} \\ \gamma^{\text{HF}}_{\mu \nu}, & \text{otherwise} \end{cases}
\label{eq:rdm1-merge}
\end{equation}

\subsection{Optimization of 1-RDM}

\begin{figure}[ht!]
    \centering
    \includegraphics[width=0.9\textwidth]{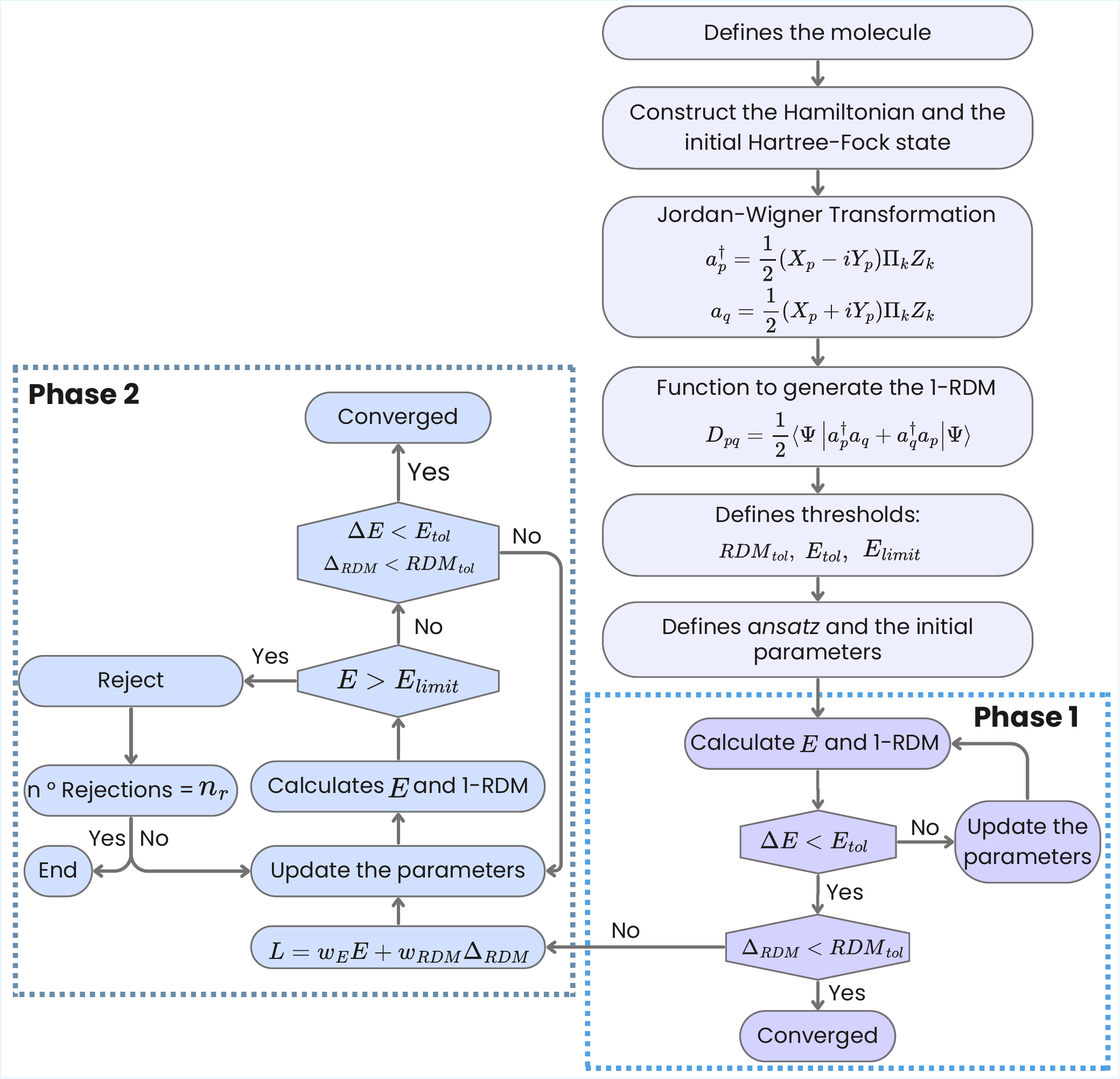}
    \\
    \caption{Flowchart for obtaining simultaneous optimization of 1-RDM and energy through VQE.}
    \label{fig:fluxograma_rdm1_conv}
\end{figure}

A code was created for the simultaneous optimization of the 1-RDM (on the molecular basis) and the energy, using VQE, aiming to improve the 1-RDM and the energy of the molecule. The code flowchart is presented in Figure \ref{fig:fluxograma_rdm1_conv}.After defining the molecular geometry, the Hamiltonian, and the initial state, the fermionic operators are mapped onto Pauli operators through the Jordan–Wigner transformation \cite{fedorov2022vqe}:

\begin{equation}
\begin{aligned}
a_p^\dagger &= \frac{1}{2} \left( X_p - i Y_p \right) \prod_{k=0}^{p-1} Z_k, \\
a_q &= \frac{1}{2} \left( X_q + i Y_q \right) \prod_{k=0}^{q-1} Z_k.
\end{aligned}
\label{eq-jw}
\end{equation}

The 1-RDM on the molecular basis, or $D_{pq}$ (equation \ref{eq-rdm1}), can be constructed from the expectation values of the creation and annihilation operators obtained from the execution of the optimized quantum circuit via VQE.  To measure $D_pq$ on the quantum device, it needs to be expressed in terms of Pauli operators through Jordan-Wigner transformation (equation \ref{eq-jw}). For example, the diagonal elements of the 1-RDM correspond to the occupations of the molecular orbitals and can be obtained by measuring the expectation value of the operator $Z$ on each qubit \cite{google2020hartree}:

\begin{equation}
\langle a_p^\dagger a_p \rangle = \frac{I - \langle Z_p \rangle}{2},
\end{equation}

where $I$ is the identity.

In turn, the expectation values of the off-diagonal operators of the 1-RDM ($p \neq q$) capture quantum coherences between distinct orbitals $p$ and $q$, that is, they describe the degree of quantum superposition between two different occupied states and their relative phases \cite{hazra2024quantifying, ibm_density_matrices, landi2018quantum}. In methods like HF, the resulting wave function is a single Slater determinant with no orbital coherence, and therefore, the 1-RDM is strictly diagonal.

When more entangled quantum states are used, like in post-HF electronic structure or quantum computing methods, the 1-RDM can exhibit significant off-diagonal elements, which may be complex-valued if the wave function includes nontrivial phase relationships between orbitals. This becomes particularly relevant in quantum simulations, where fermionic operators are mapped to qubit operators using transformations such as the Jordan–Wigner transformation. This mapping often produces Pauli strings that include the Pauli-Y operator, which contains imaginary matrix elements. As a result, the expectation values of certain observables may also become complex, requiring careful handling to ensure that the quantities being measured remain physically meaningful and real.

In practice, however, quantum devices and simulators are restricted to computing the expectation values of Hermitian operators. To ensure that only real expectation values are obtained, each operator used to reconstruct the 1-RDM must be symmetrized. This is done by combining each operator with its Hermitian adjoint (Equation \ref{eq:rdm1-real}), which guarantees real-valued outcomes consistent with physical observables \cite{google2020hartree}.

%In practice, simulators or quantum computers only calculate the expected value of real operators. Therefore, it was necessary to symmetrize the operators, that is, combine each operator with its adjoint to form a Hermitian operator, ensuring that the expected values are real, as follows \cite{google2020hartree}:

\begin{equation}
\begin{aligned}
D_{pq}  &= \frac{1}{2} \braket{ \Psi(\theta) |a_p^\dagger a_q + a_q^\dagger a_p| \Psi(\theta) }.
\end{aligned}
\label{eq:rdm1-real}
\end{equation}

In summary, the 1-RDM was obtained through the equation \ref{eq:rdm1-real} with $\Psi(\theta)$ as the VQE optimized state.

After defining these functions, the variational circuit was optimized in two distinct phases. Phase 1 corresponds to the traditional VQE optimization, aiming to minimize the system energy, which is obtained from the expected value of the Hamiltonian. The second phase refines the 1-RDM (equation \ref{eq:rdm1-real}), improving the quality of the wave function.

Throughout both phases, the RMSD \cite{chai2014root} was calculated between the 1-RDM in the current iteration ($D_{pq}^{(n)}$) in relation to the 1-RDM obtained in the previous iteration ($D_{pq}^{(n-1)}$), defined as:

\begin{equation}
\Delta_{\mathrm{RDM}} = \sqrt{\frac{1}{N^2} \sum_{p,q} \left| D_{pq}^{(n)} - D_{pq}^{(n-1)} \right|^2},
\end{equation}

where \(N\) is the dimension of the 1-RDM (number of active orbitals). In Phase 1, this metric was used solely to assess whether the 1-RDM had converged as a result of energy minimization.

In Phase 2, however, \(\Delta_{\mathrm{RDM}}\) played a more central role: it was incorporated directly into the cost function as a penalty term to guide the optimization toward a more stable and physically meaningful density matrix. The new cost function \(\mathcal{L}\) was defined as a weighted combination of the system energy and the RMSD:

\begin{equation}
    \mathcal{L} = w_{\mathrm{E}} E + w_{RDM}\Delta_{\mathrm{RDM}},
\end{equation}

where \(w_{\mathrm{E}}\) and \(w_{\mathrm{RDM}}\) are adjustable weights and $\Delta_{\mathrm{RDM}}$ is the RMSD.

The transition between phases occurs automatically: once the energy variation between two consecutive iterations falls below a predefined threshold ($\text{E}_\text{tol}$) and  \(\Delta_{\mathrm{RDM}}\) remains above the threshold $\text{RDM}_{\text{tol}}$, Phase 2 begins. The first iteration of Phase 2 uses the final parameters obtained in Phase 1. Optimization continues until both the energy and \(\Delta_{\mathrm{RDM}}\) fall below their respective thresholds, indicating full convergence of the wave function and its associated density matrix.

A criterion was also implemented to reject parameter updates that worsen the system's energy during Phase 2. Specifically, if the energy at a trial step exceeds a predefined threshold, $\text{E}_\text{limit}$, the update is rejected. If a certain number of consecutive rejections, denoted by $n_r$, occurs, the algorithm terminates the optimization.

\subsection{System of Interest and Computational Details}

To apply these ideas, the dissociation profile of the CH$_5^+$ was analyzed, a molecule whose structure has been widely discussed in the literature. In its equilibrium geometry, CH$_5^+$ exhibits the character of a complex between CH$_3$ and H$_2^+$ (CH$_3\cdots$H$_2^+$), evolving to CH$_3^+\cdots$H$_2$ during the dissociation \cite{longo2020overlap}. It is one of the simplest and, at the same time, most enigmatic carbocations, acting as a highly reactive intermediate in processes involving hydrocarbons \cite{fleming2006nature}. Furthermore, CH$_5^+$ has relevance in several contexts, such as organic reaction mechanisms \cite{olah1997kekule}, astrochemistry \cite{herbst2005chemistry}, planetary atmospheres \cite{smith1992ion}, and mass spectrometry \cite{kramer1999ch5+}.

The dissociation coordinate of CH$_5^+$ (Figure \ref{fig:ch5+structure}) was defined as the distance between the carbon and the center of mass of the exiting H$_2$, which is $1.3$ \AA\ in the equilibrium region. Thus, dissociation geometries with distances of $1.3$, $1.4$, $1.6$, $1.8$, and $2.1$ \AA\ were obtained. These distances were chosen to provide a comprehensive view of the dissociation curve of these molecules, capturing both the binding and dissociation regimes.  

Geometry optimization and classical energies were obtained using Configuration Interaction Singles and Doubles (CISD) \cite{helgaker2013molecular}. CISD is responsible for highly accurate electronic correlation effects, making it an excellent reference standard. All classical calculations were performed using Gaussian 09 \cite{frish2009gaussian}.

To minimize the computational cost of VQE calculations, in all calculations, we use the STO-3G basis set \cite{hehre1969self, collins1976self}. On this basis, each atomic orbital is represented by a linear combination of three Gaussian functions, making this the least computationally demanding basis. \cite{cramer2013essentials}. In quantum computing, using the STO-3G basis means that fewer qubits are needed to represent molecules. All quantum computer simulations were implemented using the PennyLane library \cite{bergholm2018pennylane}.

In preliminary tests, we used the \textit{ansätze} UCCSD \cite{romero2018strategies}, k-UpCCGSD \cite{lee2018generalized}, GRSD \cite{wang2024entanglement}, GateFabric, and the Entanglement-variational Hardware-efficient  \cite{wang2024entanglement}, with active spaces formed by 4, 8, 12, and 16 qubits. In these tests, the k-UpCCGSD (with $k=1$) demonstrated excellent energy accuracy compared to the CISD method when applied to the active space (4,4), i.e., 4 electrons distributed in 4 molecular orbitals, corresponding to 8 qubits. For this reason, both the k-UpCCGSD and the active space (4,4)  were selected to investigate whether the joint optimization of 1-RDM and energy allows obtaining molecular properties with good accuracy compared to CISD. The results obtained with and without 1-RDM optimization were then compared.

Although the energy from the previous test showed good agreement with CISD, a scenario in which the energy exhibits significant errors was also evaluated to test the robustness of the method in obtaining molecular properties. For this purpose, the GateFabric was used, which presented significant deviations in the CH$_5^+$ energy both in the equilibrium geometry and in the first stage of dissociation ($R = 1.4$ \AA), when applied to the active space (2,2). That is, with 2 electrons in 2 molecular orbitals (4 qubits). Thus, this second scenario (GateFabric with reduced active space) was also included in the analysis for these two geometries.

In the code developed to optimize 1-RDM in the VQE flow, the following threshold condition: $\text{E}_\text{tol}$ and $\text{RDM}_{\text{tol}}$ equal to $1\times10^{-6}$; $n_r = 10$ and \(w_{\mathrm{E}} = w_{\mathrm{RDM}} = 1\); the allowed energy worsening limit in Phase 2 was set to $E_\text{limit} = E + 1\times10^{-4}$, where $E$ is the energy of the last step of Phase 1. The optimization process in both phases used the stochastic gradient method (SGD), with the learning rate equal to $0.4$. The topological analysis of this density was performed with the Critic2 program \cite{otero2014critic2}, after generating the density file in .cube format, using the PySCF Cubegen module, with a resolution of $0.1$.

\section{Results and discussion}

\subsection{Improving Molecular Properties with 1-RDM Optimization}

Initially, the molecular properties of the CH$_5^+$ dissociation geometries were evaluated using VQE with \textit{ansatz} k-UpCCGSD and the active space (4,4), considering two scenarios: (i) 1-RDM obtained by optimizing only the energy, i.e., traditional VQE; and (ii) 1-RDM, n the molecular basis, obtained by jointly optimizing the energy and this matrix, referred to in this article as VQE$*$. The results were compared with those obtained by CISD(4,4).

The $\Delta_{\text{RDM}}$ obtained with VQE$*$ was about two decimal places smaller than that of traditional VQE. This shows that the RMSD of 1-RDM was reduced, improving the quality of the description of this matrix (Table \ref{tab:rdm1-energy}). This improvement is important because the accuracy of 1-RDM is essential to correctly describe molecular properties, especially in systems with strong electron correlation \cite{gibney2022density}. Figure \ref{fig:rdm1-dif-as4} shows the differences between the 1-RDM of the CISD and VQE methods in the active space orbitals. 

In general, the 1-RDM optimization improves the energy by $10^{-5}$ to $10^{-7}$ Hartree, and the number of additional steps varies from 7 to 39 (Table \ref{tab:rdm1-energy}). Thus, the 1-RDM optimization does not significantly improve the energies, since even before optimizing this matrix, the energy is already close to the CISD value. However, as will be shown later, this optimization can be useful when the initial wavefunction does not achieve sufficient accuracy for density-dependent properties, such as electron density.

\begin{table}[ht]
\begin{center}
\caption{Comparison between the energy (in Hartree) obtained with VQE and VQE$*$, using the \textit{ansatz} k-UpCCGSD and the active space (4,4). The energy difference between VQE$*$ and VQE (E$_\text{dif}$), the $\Delta_{\text{RDM}}$, and the number of steps are presented. The values in parentheses refer to Phase 2 of the algorithm.}
\small
\centering
\scalebox{1}{%
\begin{tabular}{ccccccc}
\hline
\textbf{R (\AA)} & \textbf{VQE} & \textbf{VQE*} & \textbf{CISD} &   \textbf{E$_\text{dif}$} & $\Delta_{\text{RDM}}$  & \textbf{Steps} \\
\hline
\textbf{1.3} &  -39.91925577  & -39.91925646   & -39.91925797 & 6.90E-7  &  6.40E-5 (9.22E-7)   &   7 (13)  \\
\textbf{1.4} & -39.9188829& -39.91888334   & -39.91888320 & 4.40E-7   &  4.48e-5 (7.64e-7)   & 7 (12) \\
\textbf{1.6} & -39.91533068 & -39.91533128  & -39.91533431 &6.00E-6   & 1.07e-5 (9.09e-7)  & 8 (7)    \\
\textbf{1.8} & -39.91710214& -39.91711269    & -39.91716922 &1.06E-5  &     5.05e-5 (9.90e-7) & 17 (15)  \\
\textbf{2.1} & -39.90742416  & -39.90742620  & -39.90742554 & 2.04E-6  &  6.96E-5 (9.55E-7)   & 14 (39)  \\
\hline
\end{tabular}%
} 
\\
\label{tab:rdm1-energy}
\end{center}
\end{table}

\subsubsection{Dipole Moment}

By optimizing 1-RDM, the errors in the dipole moments obtained with VQE with respect to the CISD reference values become significantly smaller than those obtained without this additional step (Table~\ref{tab:dipole_ch5+_as4}). This improvement is more significant at the distance of $R = 1{.}8$~\AA, where the error in the dipole moment provided by VQE without 1-RDM optimization is $-4{.}65 \times 10^{-1}$~Debye, while with the optimized 1-RDM this error is reduced to $-7{.}00 \times 10^{-5}$~Debye, a difference of approximately four orders of magnitude. These results demonstrate that the optimization of 1-RDM contributes to the improvement in the description of the dipole moment along the CH$_5^+$ dissociation.

\begin{table}[H]
\begin{center}
\caption{Total dipole moment (in Debye) and differences obtained by means of VQE (k-UpCCGSD) and CISD for CH$_5^+$ in different dissociation geometries, in the active space (4,4).}
\small
\centering
\label{tab:dipole_ch5+_as4}
\scalebox{1}{%
\begin{tabular}{crrrrr}
\hline
\textbf{R (\AA)} & \textbf{VQE} & \textbf{VQE*}  & \textbf{CISD} & \textbf{Error} & \textbf{Error*} \\
\hline
1.3   & 1.9652 &  1.9212     & 1.9210       &   4.42E-2 &  1.60E-4    \\
1.4   & 1.8170  &   1.8140    & 1.8149       & 2.12E-3   &  -9.10E-4   \\
1.6   & 1.1196 &  1.1190     & 1.1189       &  6.90E-4 &    9.00E-5 \\
1.8   & 0.1142 &   0.5787    & 0.5788       & -4.65E-1 &   -7.00E-5 \\
2.1   & 0.6913  &    0.6769   & 0.6785        & 1.29E-2   &  -1.59E-3   \\ 
\hline
\end{tabular}%
} 
\\
\end{center}
\end{table}

\subsubsection{Electron Density}

\begin{figure}[ht!]
\centering
\caption{Position of  CH$_5^+$ atoms, obtained with CISD/STO-3G and $R = 1.3$ \AA.}
\includegraphics[width=0.25\textwidth]{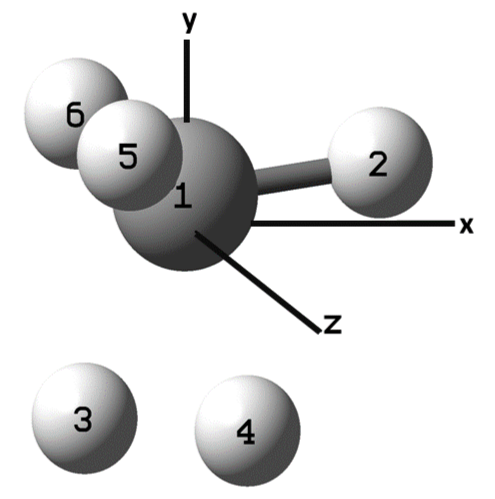}
\label{fig:mol}
\\
\end{figure}

\begin{figure}[ht!]
\centering
\caption{Difference in electron density (in e/Bohr$^3$) along the \(x\) and \(y\) axes obtained with VQE (\textit{ansatz} k-UpCCGSD) and CISD for the CH$_5^+$ dissociation geometries, using the STO-3G basis and the active space (4,4).}
\includegraphics[width=0.77\textwidth]{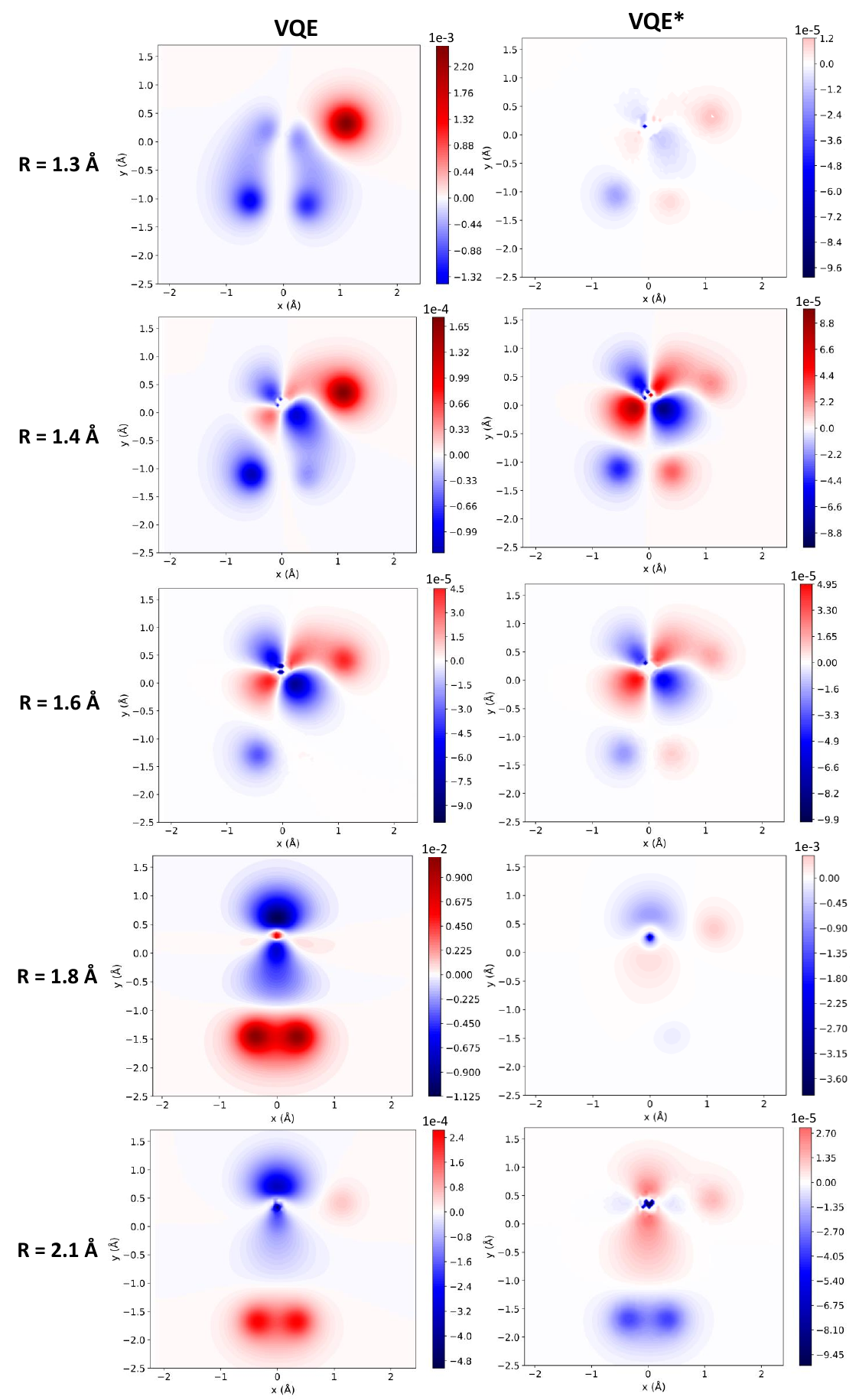}
\label{fig:elec-density-dif}
\\
\end{figure}

Figure \ref{fig:elec-density-dif} shows the difference in electron density between the VQE (with and without 1-RDM optimization) and CISD methods for different CH$_5^+$ dissociation geometries. The red and blue regions indicate higher electron density in the VQE and CISD calculations, respectively.  For spatial reference, Figure \ref{fig:mol} shows the atomic configuration of CH$_5^+$ at equilibrium geometry, which corresponds to the same molecular structure used in comparisons of Figure \ref{fig:elec-density-dif}.

Figure~\ref{fig:elec-density-dif} presents the electron density differences between the VQE results (with and without 1-RDM optimization) and the CISD reference for various CH$_5^+$ dissociation geometries. In these plots, red and blue regions indicate higher electron density in the VQE and CISD calculations, respectively. For spatial reference, Figure~\ref{fig:mol} shows the atomic configuration of CH$_5^+$ at equilibrium geometry, which corresponds to the same molecular structure used in the electron density comparisons of Figure~\ref{fig:elec-density-dif}.

In general, the 1-RDM optimization results in smoother electron density difference maps with reduced error scales (ranging from $10^{-3}$ to $10^{-5}$) compared to the unoptimized 1-RDM, which presents errors between $10^{-2}$ and $10^{-5}$. The main exception occurs at $R = 1.6$ \AA, where the maps with and without 1-RDM optimization are similar, but whose VQE* gives slightly better results. This suggests that, for this configuration, the wave function already had a sufficiently high quality before the optimization of this matrix was performed. Nevertheless, for the other geometries, it is observed that even when the ground state energy approaches the exact value, the quality of the electron density can be improved through the optimization of the 1-RDM. This improvement may be important for local properties or properties that are highly sensitive to the electron distribution.

These results are corroborated by the topological analysis of the electron density, that is, of the values of the density and its Laplacian ($\nabla^2 \rho$) at the nuclear critical points (NCPs) and bonding critical points (BCPs) (Table \ref{Tab:ch5+electron_density_as4}). Figure \ref{fig:bcp_critic2} shows the location of the BCPs throughout the dissociation, highlighting that even in highly dissociated geometries ($1.8$ and $2.1$ Å), there are still critical points between the carbon and the distant hydrogens. VQE was able to identify the same NCPs and BCPs as CISD, demonstrating that the electron density topology was correctly preserved, even with an approximate variational \textit{ansatz} and dissociated geometries. This indicates that VQE can correctly capture weak interactions between fragments, a common challenge for approximate methods.

Small discrepancies in the absolute values of $\rho$ and $\nabla^2 \rho$ at the critical points indicate that VQE produces slightly different results, with errors ranging from $10^{-8}$ to $10^{-3}$. These deviations, however, are small and confirm the robustness of VQE in describing the electron density and its topological aspects. It is observed, however, that the density Laplacian ($\nabla^2 \rho$) presents more significant errors, especially in nuclear regions (Table \ref{Tab:ch5+electron_density_as4}). In the carbon nucleus, for example, the deviations can reach the order of $10^2$, which is due to the high concentration of density in this region and the numerical sensitivity resulting from very high values (Table \ref{tab:density-SI}).

In general, the optimization of the 1-RDM contributes to the reduction of errors in the electron density and in the $\nabla^2 \rho$ obtained with the VQE. For example, a clear reduction in the error of the $\nabla^2 \rho$ of the carbon NCP is observed at $R=1.6$ \AA, since without the optimization of the 1-RDM the error obtained was $1{.}48$, whereas when performing the optimization, the error is reduced to $-0{.}70$. Furthermore, the errors in the $\nabla^2 \rho$ of the BCP decrease by one to two decimal places when optimizing this matrix. The improvement in the representation of $\nabla^2 \rho$ of the BCP has direct implications in the characterization of the type of chemical interaction involved, since negative values are characteristic of covalent bonds \cite{cortes2023introduction}. Therefore, reducing the error at these critical points increases confidence in the topological classification of bonds and in the physicochemical interpretation of the electronic structure.

\begin{table}[ht]
\begin{center}
\caption{Difference between the results of the electron density (in e/Bohr$^3$) and the Laplacian of the critical points of the nuclei (NCP) and the bonds (BCP) in the dissociation geometries of CH$_{5}^{+}$, which were obtained with the VQE and VQE* compared to CISD, using the active space (4,4) and the STO-3G basis.}
\small
\centering
\label{Tab:ch5+electron_density_as4}
\scalebox{1}{%
\begin{tabular}{ccccccccccc}
\hline
\textbf{CP} & \textbf{1.3} & \textbf{1.3*} & \textbf{1.4} & \textbf{1.4*} & \textbf{1.6} & \textbf{1.6*} & \textbf{1.8} & \textbf{1.8*} & \textbf{2.1} & \textbf{2.1*} \\ \hline
\multicolumn{11}{c}{\textbf{Electron density}}                                                                                                                                         \\ \hline
C1  & -4.10E-5                       & 3.00E-5& 4.40E-5                        & 1.51E-4                         & -4.90E-5                       & 6.00E-6                         & 9.97E-3                        & -4.93E-3                        & -9.97E-4                       & -1.18E-4                        \\
H2 & 2.67E-3                        & 1.27E-5& 1.85E-4                        & 2.35E-5                         & 4.52E-5                        & 1.69E-5                         & 6.73E-5                        & 4.09E-4                         & 5.91E-5                        & 1.53E-5                         \\
H3 & -1.45E-3                       & -1.80E-5& -1.32E-4                       & -4.85E-5                        & -3.49E-5                       & -2.05E-5                        & 1.15E-2                        & 5.49E-6                         & 2.79E-4                        & -3.94E-5                        \\
H4 & -1.15E-3                       & 7.48E-6& -3.47E-5                       & 3.34E-5                         & 1.94E-6                        & 1.03E-5                         & 1.14E-2                        & -1.41E-4                        & 2.74E-4                        & -4.01E-5                        \\
H5 & 6.49E-4                        & 3.14E-6& 3.94E-5                        & -2.16E-6                        & 1.01E-5                        & 7.00E-8                         & 7.71E-5                        & 3.83E-4                         & 5.84E-5                        & 1.31E-5                         \\
H6 & 6.49E-4                        & 3.14E-6& 3.94E-5                        & -2.16E-6                        & 1.01E-5                        & 7.10E-8                         & 7.71E-5                        & 3.83E-4                         & 5.84E-5                        & 1.31E-5                         \\
C1-H5  & 1.75E-4                        & 1.08E-6& 1.19E-5                        & -6.58E-7                        & 5.08E-6                        & -4.00E-9                        & -4.55E-5                       & 3.61E-5                         & 1.67E-4                        & 1.12E-6                         \\
C1-H6   & 1.75E-4                        & 1.08E-6& 1.19E-5                        & -6.58E-7                        & 5.03E-6                        & -5.80E-8                        & -4.55E-5                       & 3.61E-5                         & 8.53E-6                        & 1.12E-6                         \\
C1-H2  & 7.23E-4                        & 4.01E-6& 4.68E-5                        & 3.83E-6                         & 1.24E-5                        & 3.96E-6                         & -4.50E-5                       & 3.91E-5                         & -1.51E-4                       & 2.05E-6                         \\
H3-H4  & -4.33E-7                       & -3.80E-8& -1.80E-6                       & -1.04E-6                        & -3.37E-6                       & -1.43E-6                        & 6.61E-3                        & -3.39E-5                        & 1.74E-4                        & -2.49E-5                        \\
C1-H3  & -4.15E-5                       & 9.42E-7& 1.59E-6                        & 2.50E-6                         & -2.88E-6                       & 8.25E-7                         & -2.96E-3                       & 9.97E-5                         & -3.91E-5                       & 9.52E-6                         \\ \hline
\multicolumn{11}{c}{\textbf{Laplacian}}                                                                                                                                                                                                                                                                                                                                       \\ \hline
C1 & 1.37                         & -1.07                         & -1.50                      & -5.16                      & 1.48                        & -1.70E-1                        & -3.32E+2                       & 1.64E+2                         & 3.35                      & 3.82                          \\
H2  & 1.82E-1                        & 6.78E-4                         & 1.16E-2                        & 1.31E-3                         & 3.00E-3                        & 1.34E-3                         & 2.47E-3                        & 2.25E-2                         & 2.49E-3                        & -6.83E-5                        \\
H3  & -1.04E-1                       & -1.40E-3                        & -5.16E-3                       & -2.12E-3                        & -2.62E-3                       & -1.52E-3                        & 5.14E-1                        & 1.66E-4                         & 9.06E-3                        & -1.08E-3                        \\
H4  & -5.24E-2                       & -1.35E-4                        & -2.86E-3                       & 1.29E-3                         & -5.18E-4                       & 2.88E-4                         & 4.27E-1                        & -4.43E-3                        & 1.11E-2                        & -1.33E-3                        \\
H5  & 3.00E-2                        & 3.04E-4                         & 3.85E-3                        & 2.40E-5                         & 8.59E-4                        & -1.08E-4                        & 1.50E-3                        & 1.37E-2                         & 7.88E-3                        & 1.69E-3                         \\
H6  & 3.00E-2                        & 3.04E-4                         & 3.85E-3                        & 2.39E-5                         & 8.59E-4                        & -1.08E-4                        & 1.50E-3                        & 1.37E-2                         & 7.87E-3                        & 1.69E-3                         \\
C1-H5  & 1.61E-3                        & -3.55E-5                        & -1.68E-4                       & 5.08E-4                         & 1.19E-4                        & -1.64E-4                        & 1.24E-2                        & 2.30E-3                         & -4.90E-2                       & 6.08E-4                         \\
C1-H6   & 1.60E-3                        & -3.53E-5                        & -1.68E-4                       & 5.08E-4                          & 1.20E-4                        & -1.64E-4                        & 1.21E-2                        & 1.69E-3                         & 2.20E-3                        & 5.63E-4                         \\
C1-H2  & 9.82E-4                        & -1.37E-3                        & 3.37E-4                        & -1.38E-4                          & 1.75E-3                        & 1.64E-3                         & -3.99E-4                       & 1.55E-3                         & 5.11E-2                        & 2.42E-4                         \\
H3-H4  & -3.40E-3                       & 1.10E-4                         & -4.79E-5                       & 1.05E-4                      & -1.06E-4                       & 2.94E-4                         & -1.89E-2                       & 6.62E-4                         & -2.08E-4                       & -3.34E-5                        \\
C1-H3  & 1.98E-3                        & -2.80E-4                        & -9.84E-4                       & 9.43E-5                         & 2.70E-6                        & -1.09E-5                        & 1.66E-2                        & -2.61E-4                        & 1.63E-4                        & 2.16E-4                         \\ \hline
\end{tabular}%
} 
\\
\end{center}
\end{table}

\subsubsection{Electrostatic Potential}

\begin{figure}[ht!]
\centering
\includegraphics[width=0.7\textwidth]{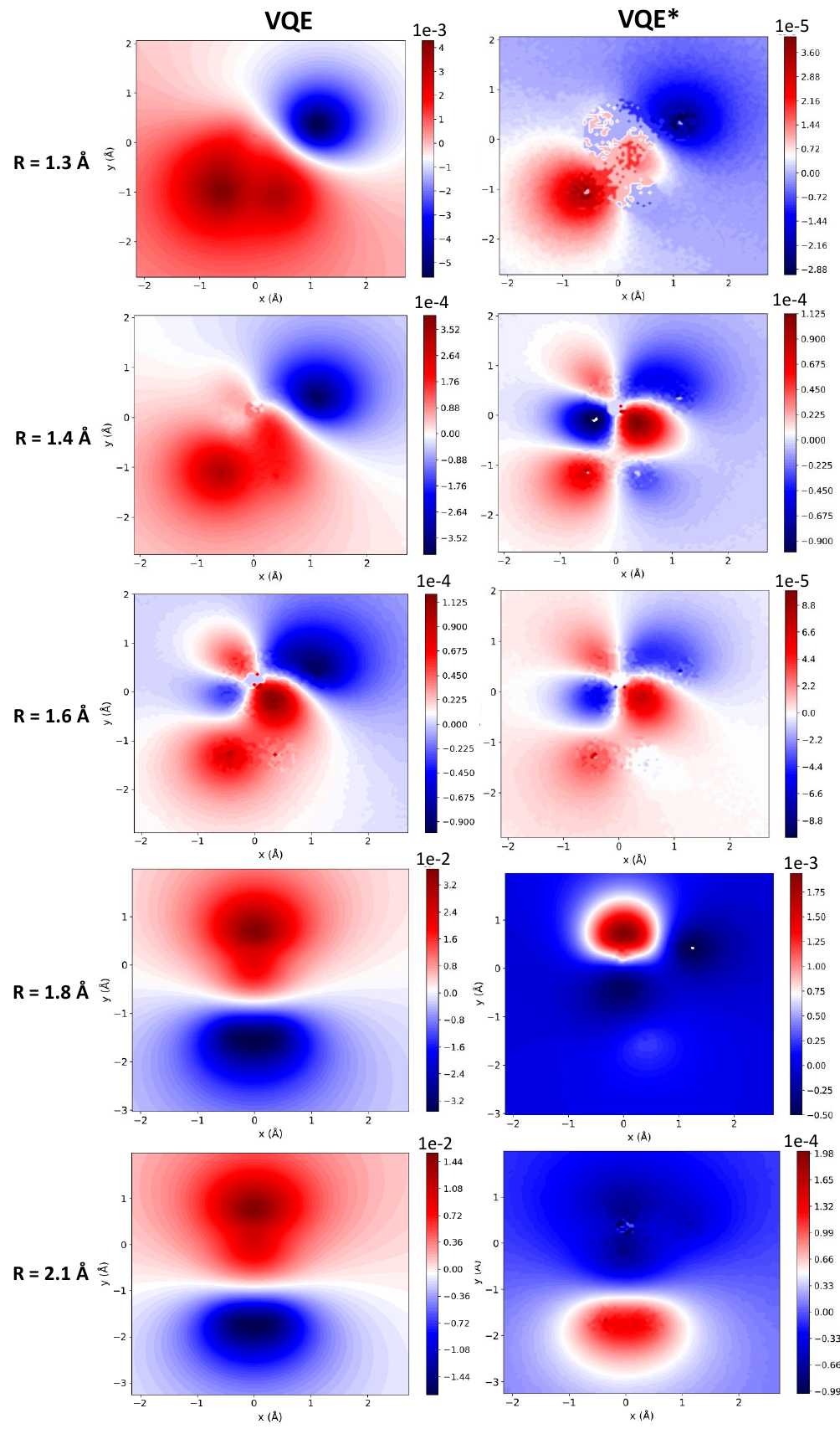}
\\
\caption{Difference in electrostatic potential (in e/Bohr$^3$) along the \(x\) and \(y\) axes, in the active space (4,4), between the VQE and VQE* compared to CISD, along the CH$_5^+$ dissociation geometries, using the STO-3G basis.}
\label{fig:dif-pot-elec}
\end{figure}

The analysis of the electrostatic potential complements the discussions on the electron density, since this property is obtained from the electron density (equation \ref{eq-potencial_elet}). Consequently, the differences in the electrostatic potential between VQE and VQE* compared to CISD (Figure \ref{fig:dif-pot-elec}) follow trends similar to those observed in the electron density variations. Consequently, the errors in the electrostatic potential generated with the optimized 1-RDM are smoother and present smaller absolute values, especially in poorly dissociated geometries, as indicated by the color scale (Figure \ref{fig:dif-pot-elec}). Therefore, the optimization of the 1-RDM contributes significantly to the improvement of the electrostatic potential approximation.

\subsubsection{Mulliken Population Analysis}

Table \ref{tab:mulliken_pop_ch5+_as4} presents the differences in Mulliken charges and populations of CH$_5^+$ atoms along different dissociation geometries, obtained by VQE$*$ and CISD. In general, differences after the second decimal place in the population analysis obtained by different methods are not considered significant \cite{kang2008efficient, lu2006atomic}. In this work, the absolute values of these differences are mostly between $10^{-2}$ and $10^{-5}$, which indicates good agreement between VQE and CISD (Table \ref{tab:mulliken_pop_ch5+_as4}). In general, the errors in the Mulliken loadings and populations obtained with VQE are one to two decimal places smaller when optimizing the 1-RDM compared to the traditional VQE, which optimizes only the energy. This reinforces the positive impact of optimizing this matrix on the accuracy of the Mulliken population analysis.

\begin{table}[ht]
\begin{center}
\small
\centering
\caption{Difference in Mulliken charges and populations (in atomic charge units) obtained by VQE (k-UpCCGSD) and CISD for the CH$_5^+$ dissociation geometries, from 1.3 to 2.1 \AA, using the active space (4,4) and the STO-3G basis. * indicates the results obtained with the optimized 1-RDM.}
\label{tab:mulliken_pop_ch5+_as4}
\scalebox{1}{%
\begin{tabular}{ccccccccccc}
\hline
\textbf{Átomo} & \textbf{1.3} & \textbf{1.3*} & \textbf{1.4} & \textbf{1.4*} & \textbf{1.6} & \textbf{1.6*} & \textbf{1.8} & \textbf{1.8*} & \textbf{2.1} & \textbf{2.1*} \\
\hline
\multicolumn{11}{c}{\textbf{Mulliken Charges}} \\ \hline
C1                      & 2.58E-3                         & 2.00E-5      & 1.90E-4     & 1.00E-5      & 8.00E-5     & 2.00E-5      & 5.09E-2     & 2.11E-3      & 1.52E-3      & -1.10E-4      \\
H2                     & -5.82E-3                        & -3.00E-5     & -4.00E-4    & -5.00E-5     & -1.10E-4    & -4.00E-5     & -8.00E-5    & -8.70E-4     & -1E-4       & -2.00E-5        \\
H3                     & 3.41E-3                         & 4.00E-5      & 3.00E-4     & 1.00E-4      & 8.00E-5     & 4.00E-5      & -2.55E-2    & 1.00E-5      & -6.00E-4      & 9.00E-5         \\
H4                     & 2.67E-3                         & -1.00E-5     & 1.00E-4     & -6.00E-5     & 0.00      & -2.00E-5     & -2.51E-2    & 3.60E-4      & -5.90E-4     & 9.00E-5         \\
H5                     & -1.41E-3                        & -1.00E-5     & -8.00E-5    & 1.00E-5      & -2.00E-5    & 0.00       & -1.10E-4    & -8.10E-4     & -1.10E-4     & -2.00E-5        \\
H6                     & -1.41E-3                        & -1.00E-5     & -8.00E-5    & 1.00E-5      & -2.00E-5    & 0.00       & -1.10E-4    & -8.10E-4     & -1.10E-4     & -2.00E-5        \\ \hline
\multicolumn{11}{c}{\textbf{Mulliken population}} \\ \hline
C1 (1s)                 & 0.00                          & 0.00       & 0.00      & 0.00       & 0.00      & 0.00       & 2.00E-5     & -1.00E-5     & 0            & 0             \\
C1 (2s)                 & 0.00                          & 0.00       & 1.00E-5     & 1.00E-5      & -1.00E-5    & 0.00       & 2.08E-3     & -1.24E-3     & -1.80E-4    & -5.00E-5        \\
C1 (2px)                & -2.50E-3                        & -1.00E-5     & -1.60E-4    & 1.00E-5      & -2.00E-5    & -1.00E-5     & 1.00E-5     & 0.00       & 0.00              &0.00               \\
C1 (2py)                & -9.00E-5                        & -1.00E-5     & -5.00E-5    & -3.00E-5     & -7.00E-5    & 0.00       & -5.30E-2    & -8.40E-4     & -1.35E-3     & 1.70E-4       \\
C1 (2pz) & 1.00E-5                         & 0.00       & 1.00E-5     & 0.00       & 3.00E-5     & 0.00       & -2.00E-5    & -2.00E-5     & 1.00E-5        & -1.00E-5        \\
H2 (1s)                & 5.82E-3                         & 3.00E-5      & 4.00E-4     & 5.00E-5      & 1.10E-4     & 4.00E-5      & 8.00E-5     & 8.70E-4      & 1.00E-4        & 2.00E-5         \\
H3 (1s)                & -3.41E-3                        & -4.00E-5     & -3.00E-4    & -1.00E-4     & -8.00E-5    & -4.00E-5     & 2.55E-2     & -1.00E-5     & 6.00E-4       & -9.00E-5        \\
H4 (1s)                & -2.67E-3                        & 1.00E-5      & -1.00E-4    & 6.00E-5      & 0.00      & 2.00E-5      & 2.51E-2     & -3.60E-4     & 5.90E-4      & -9.00E-5        \\
H5 (1s)                & 1.41E-3                         & 1.00E-5      & 8.00E-5     & -1.00E-5     & 2.00E-5     & 0.00       & 1.10E-4     & 8.10E-4      & 1.10E-4      & 2.00E-5         \\
H6 (1s)                & 1.41E-3                         & 1.00E-5      & 8.00E-5     & -1.00E-5     & 2.00E-5     & 0.00       & 1.10E-4     & 8.10E-4      & 1.10E-4      & 2.00E-5         \\ \hline

\end{tabular}%
} 
\\
\end{center}
\end{table}

\subsection{Improving Energy with 1-RDM Optimization}

In this section, we compare the results obtained using VQE, with and without 1-RDM optimization, employing the \textit{ansatz} GateFabric and the active space (2,2) for $R = 1.3$ and $1.4$ \AA. The goal is to evaluate whether the 1-RDM optimization is capable of improving the description of the CH$_5^+$ energy, in addition to the derived properties presented previously. 

The choice of the GateFabric \textit{ansatz} and the active space (2,2) is justified by the fact that, without 1-RDM optimization, the VQE results presented significant discrepancies in relation to the energy obtained via CISD in the same active space, with errors of $0.25$ and $0.14$ Hartree (Table \ref{tab:rdm1-energy-gf}). With VQE*, a variation of $0.26$ and $0.14$ Hartree was observed in relation to the values obtained without optimization, resulting in a reduction of the error compared to CISD to only $10^{-8}$ Hartree (Figure \ref{tab:rdm1-energy-gf}). 

\begin{table}[ht]
\begin{center}
\caption{Comparison between the energy (in Hartree) obtained with VQE and VQE$*$, using the \textit{ansatz} GateFabric and the active space (2,2). The energy difference between VQE$*$ and VQE (E$_\text{dif}$), the $\Delta_{\text{RDM}}$, and the number of steps are presented. The values in parentheses refer to Phase 2 of the algorithm.}
\small
\centering
\scalebox{1}{%
\begin{tabular}{ccccccc}
\hline
\textbf{R (\AA)} & \textbf{VQE} & \textbf{VQE*} & \textbf{CISD} &   \textbf{E$_\text{dif}$} & $\Delta_{\text{RDM}}$  & \textbf{Steps} \\
\hline
\textbf{1.3} &  -39.65778932  &-39.91758947   & -39.91758946 & 0.2598  &  1.51E-2 (9.36e-7) &   23 (11)  \\
\textbf{1.4} & -39.77277307 & -39.91761935  & -39.91761933 & 0.1448    &   1.33E-2 (7.20e-7)  & 41 (12) \\
\hline
\end{tabular}%
} 
\\
\label{tab:rdm1-energy-gf}
\end{center}
\end{table}

\begin{figure}[ht!]
\centering
\caption{Difference in electron density (in e/Bohr$^3$) along the \(x\) and \(y\) axes obtained with VQE and VQE* compared to CISD, for the CH$_5^+$ dissociation geometries, using the STO-3G basis and the active space (2,2).}
\includegraphics[width=0.8\textwidth]{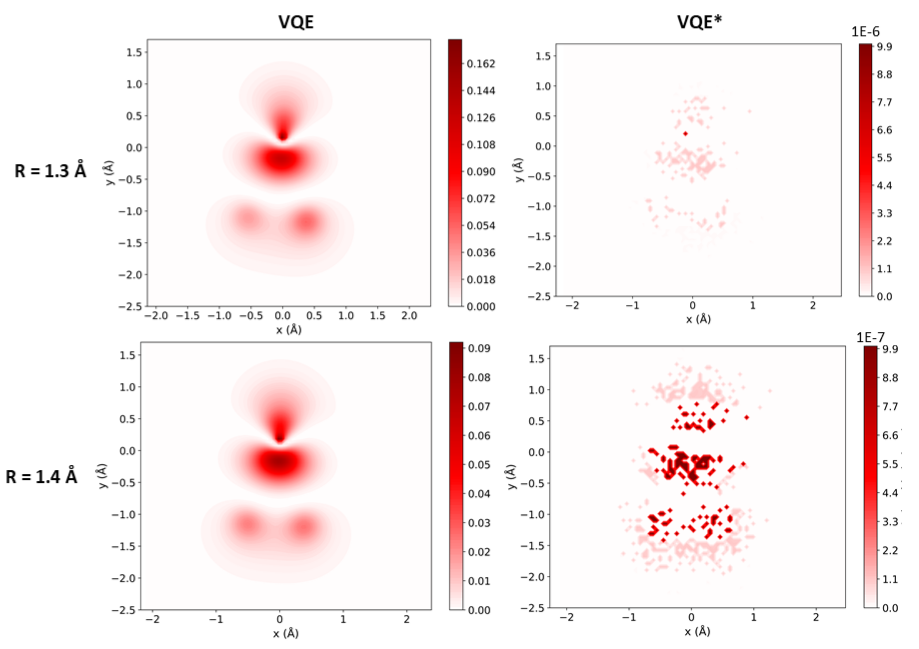}
\label{fig:elec-density-dif-gf}
\\
\end{figure}

The 1-RDM generated by VQE presented significant errors, whose $\Delta_{\text{RDM}}$ was approximately $10^{-2}$, while in VQE* this error was reduced to $10^{-7}$ (Table \ref{tab:rdm1-energy-gf}). This shows a significant improvement in the description of the 1-RDM when using VQE*. This can also be observed in the differences between the 1-RDM generated by VQE and VQE* in relation to CISD, presented in Figure \ref{fig:rdm1-dif-gf}. This improvement in the quality of the 1-RDM was also reflected in the description of the electron density, as evidenced in Figure \ref{fig:elec-density-dif-gf}, indicating a much more accurate representation of the system.

\section{Conclusion}

%In this work, we evaluate the molecular properties of the CH$_5^+$ dissociation geometries using the traditional VQE algorithm (energy-only optimization) and a VQE approach with self-consistent and joint optimization of the energy and 1-RDM. This self-consistent optimization scheme was inspired by classical computational chemistry methods.

In this work, we propose and validate a modified Variational Quantum Eigensolver (VQE) that incorporates the optimization of the one-particle reduced density matrix (1-RDM) alongside energy minimization. Inspired by strategies commonly used in classical electronic structure methods, this approach aims to enhance not only the accuracy of ground-state energies but also of molecular properties that are directly dependent on the 1-RDM, such as dipole moments, electron densities, electrostatic potentials, and Mulliken charges and populations.

%The \textit{ansatz} k-UpCCGSD and GateFabric were used, with active spaces (4,4) and (2,2), respectively. The results were compared with CISD calculations performed on the same active spaces, aiming to validate the accuracy of the approaches.

To evaluate the effectiveness of the proposed strategy, we used the CH$_5^+$ molecule as a benchmark system due to its known electron correlation complexity and relevance in various chemical contexts. Two ansätze were employed: k-UpCCGSD with active spaces (4,4), which is chemistry-inspired and accurate for energy, and GateFabric with active spaces (2,2), which is hardware-efficient but prone to greater approximation errors. The results were compared with CISD calculations performed on the same active spaces, aiming to validate the accuracy of the approaches.

In the case of k-UpCCGSD, the inclusion of the 1-RDM optimization did not result in significant improvements in the energy of the CH$_5^+$ dissociation geometries (variations in the order of $10^{-5}$ to $10^{-7}$ Hartree). This is because the energy obtained by traditional VQE already showed excellent agreement with CISD values. However, when compared with traditional VQE, the additional optimization step of 1-RDM provided a substantial improvement in the description of molecular properties, such as dipole moment, electron density, electrostatic potential, and Mulliken charges and populations.

On the other hand, using GateFabric with traditional VQE, a considerable divergence in energy was observed in relation to CISD. However, the joint optimization of 1-RDM during the VQE process resulted in a significant improvement in both energy accuracy and the derived molecular properties.

Overall, our findings demonstrate that energy alone is not a sufficient convergence criterion for VQE when the accurate prediction of molecular properties is desired. Although VQE can yield energies comparable to those from CISD, there is no guarantee that the corresponding 1-RDM is properly optimized. Our study shows that a two-phase optimization of both the energy and the 1-RDM within the VQE framework enhances the fidelity of the wavefunction and significantly improves the accuracy of properties such as dipole moments and Mulliken charges. This procedure can also improve the energy itself, particularly when the chosen \textit{ansatz} lacks sufficient expressiveness to describe the system accurately. The proposed method thus offers a conceptually simple and computationally feasible enhancement to VQE, making it especially valuable for NISQ-era simulations where ansätze are often shallow and noise-prone. Looking ahead, this strategy could be extended to larger systems, integrated with 2-RDM refinement, or applied in multi-reference regimes, further advancing the reliability of quantum algorithms for molecular characterization.

\section*{Acknowledgments}

AM Lima acknowledges the fellowship granted by the National Council for Scientific and Technological Development (CNPq). ES Teixeira acknowledges the support provided by Venturus and its Quantum Computing Center.

%Bibliography
%\bibliographystyle{unsrt}  
%\bibliography{references}  

\clearpage
\pagebreak

\begin{center}
\textbf{\large Supplementary Materials}
\end{center}
\setcounter{equation}{0}
\setcounter{figure}{0}
\setcounter{table}{0}
\setcounter{page}{1}
\makeatletter
\renewcommand{\theequation}{S\arabic{equation}}
\renewcommand{\thefigure}{S\arabic{figure}}
\renewcommand{\thetable}{S\arabic{table}}

\begin{figure}[ht!]
    \centering
    \includegraphics[width=0.53 \textwidth]{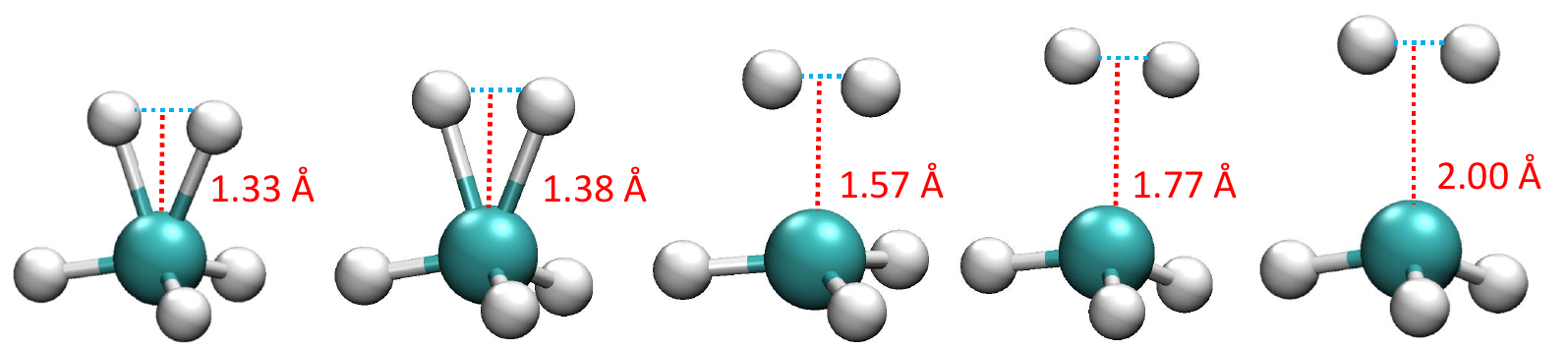}
    \caption{Dissociation intermediate geometries of CH$_5^+$.}
    \phantomsection
    \label{fig:ch5+structure}
\end{figure}

\begin{figure}[H]
    \centering
    \includegraphics[width=0.6\textwidth]{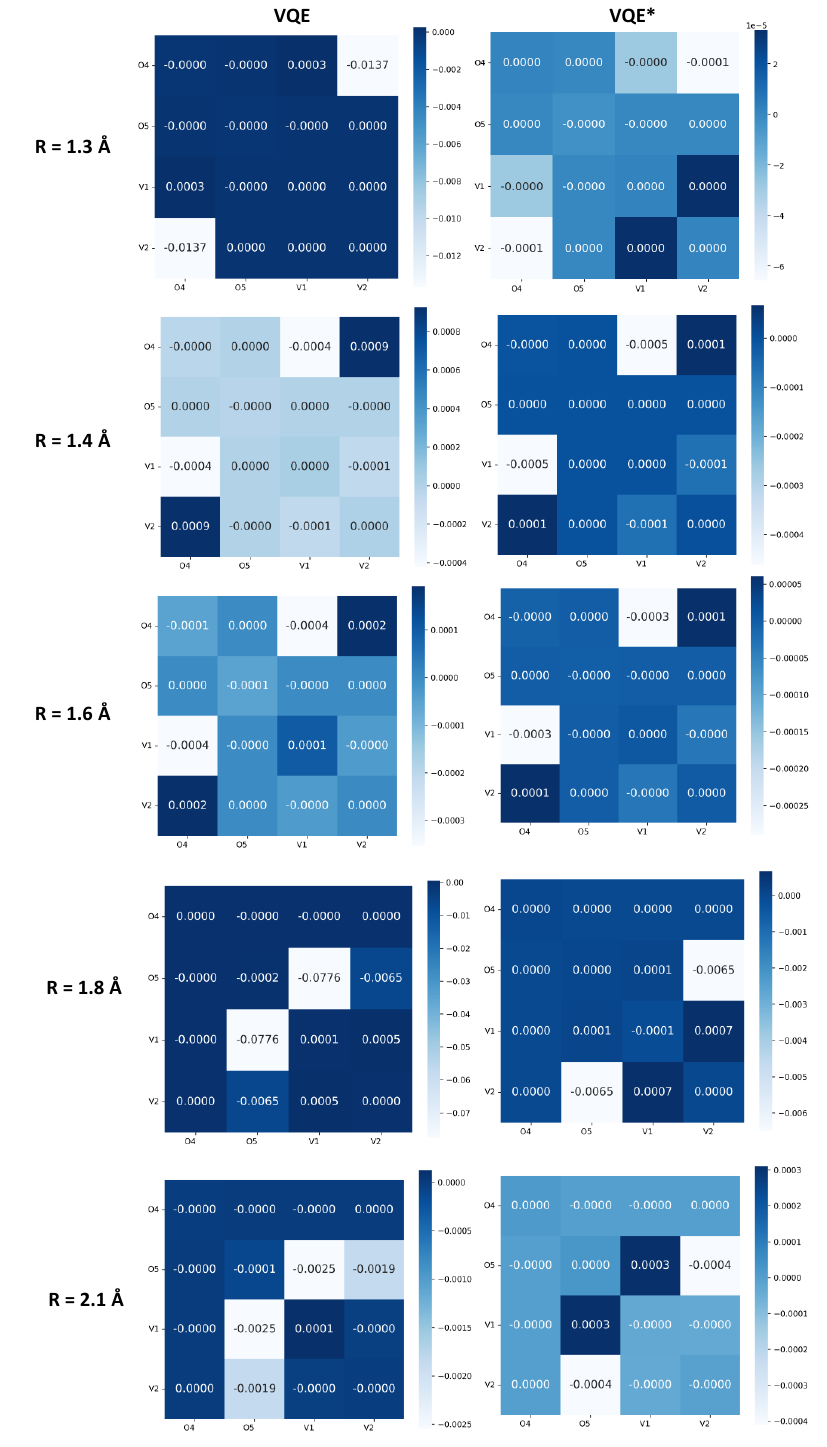}
    \caption{Difference in 1-RDM obtained with VQE and VQE* compared to CISD. The k-UpCCGSD \textit{ansatz}, the active space (4,4) and the STO-3G basis were used. O4 and O5 refer to the last two occupied orbitals, while V1 and V2 are the first two virtual orbitals.}
    \phantomsection
    \label{fig:rdm1-dif-as4}

\end{figure}

\begin{figure}[H]
    \centering      
    \includegraphics[width=1.0\textwidth]{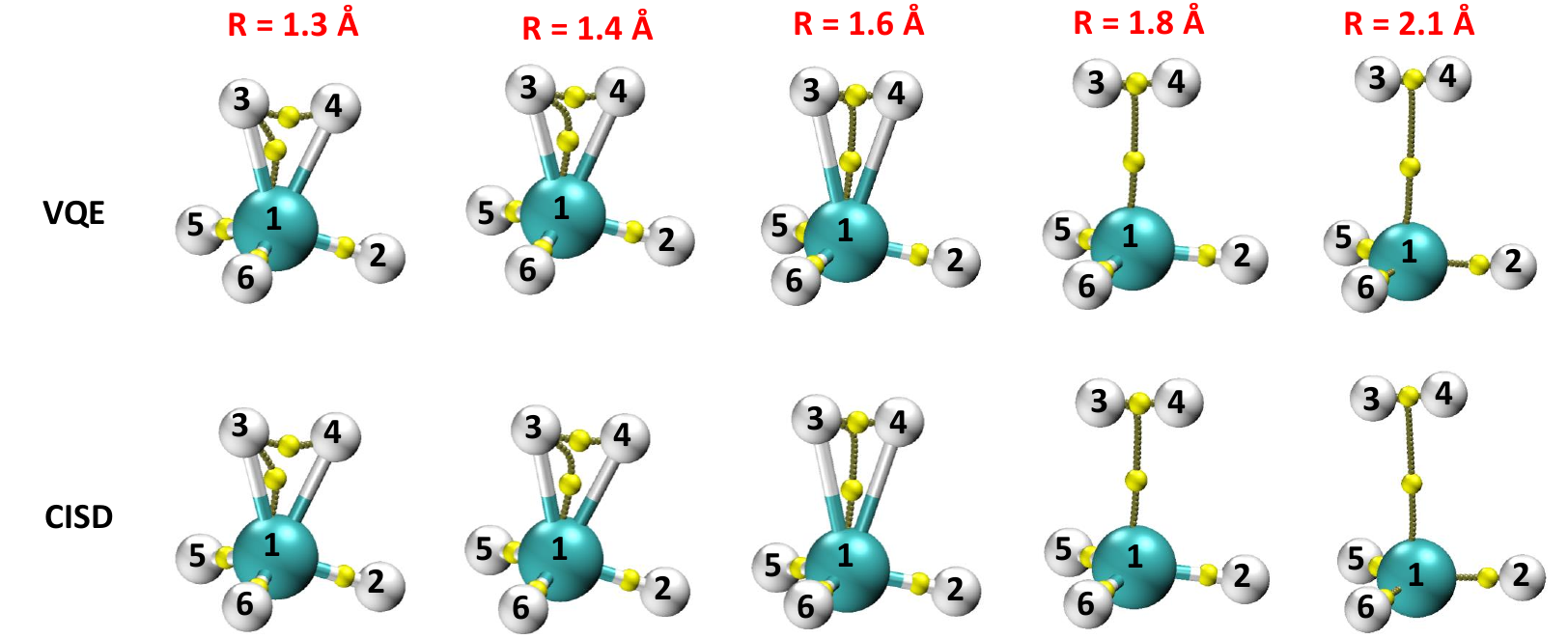}
    \caption{Critical points of the bonds in the CH$_5^+$ dissociation geometries, obtained using the CISD method and with the VQE without -1RDM optimization (active space (4,4), and \textit{ansatz} k-UpCCGSD), both with the STO-3G basis. The results with VQE* are similar to those of VQE.}
    \phantomsection
    \label{fig:bcp_critic2}
\end{figure}

\begin{table}[H]
\centering
\caption{Electron density and Laplacian of the CH$_5^+$ dissociation geometries with VQE and VQE* compared to CISD. The k-UpCCGSD, active space (4,4), and STO-3G basis were used.}
\scalebox{0.52}{
\begin{tabular}{crrrrrrrrrrrrrrr}
\hline
\textbf{CP} & \multicolumn{3}{c}{\textbf{1.3}} & \multicolumn{3}{c}{\textbf{1.4}} & \multicolumn{3}{c}{\textbf{1.6}} & \multicolumn{3}{c}{\textbf{1.8}} & \multicolumn{3}{c}{\textbf{2.1}} \\ \hline
& VQE & VQE* & CISD & VQE & VQE* & CISD & VQE & VQE* & CISD & VQE & VQE* & CISD & VQE & VQE* & CISD \\
\hline
\multicolumn{16}{c}{\textbf{Electron Density}} \\ \hline
nucleus (C) & 101.66156 & 101.66163 & 101.66160 & 121.57699 & 121.57709 & 121.57694 & 118.18339 & 118.18344 & 118.18344 & 96.03156 & 96.01666 & 96.02158 & 127.04321 & 127.04408 & 127.04420 \\
nucleus (H) & 0.30363 & 0.30097 & 0.30096 & 0.30515 & 0.30499 & 0.30496 & 0.30841 & 0.30838 & 0.30836 & 0.30366 & 0.30401 & 0.30360 & 0.29917 & 0.29912 & 0.29911 \\
nucleus (H) & 0.26071 & 0.26214 & 0.26216 & 0.27308 & 0.27317 & 0.27322 & 0.29060 & 0.29061 & 0.29063 & 0.33177 & 0.32028 & 0.32028 & 0.35587 & 0.35555 & 0.35559 \\
nucleus (H) & 0.26313 & 0.26429 & 0.26428 & 0.26856 & 0.26862 & 0.26859 & 0.29965 & 0.29965 & 0.29964 & 0.33378 & 0.32221 & 0.32235 & 0.35352 & 0.35321 & 0.35325 \\
nucleus (H) & 0.31639 & 0.31574 & 0.31574 & 0.30988 & 0.30983 & 0.30984 & 0.30512 & 0.30511 & 0.30511 & 0.30793 & 0.30823 & 0.30785 & 0.28289 & 0.28285 & 0.28284 \\
nucleus (H) & 0.31639 & 0.31575 & 0.31574 & 0.30988 & 0.30983 & 0.30984 & 0.30512 & 0.30511 & 0.30511 & 0.30793 & 0.30823 & 0.30785 & 0.28290 & 0.28286 & 0.28285 \\
Bond(C1-H5) & 0.25337 & 0.25319 & 0.25319 & 0.25300 & 0.25299 & 0.25299 & 0.25056 & 0.25056 & 0.25056 & 0.24720 & 0.24728 & 0.24724 & 0.24354 & 0.24337 & 0.24337 \\
Bond(C1-H6) & 0.25337 & 0.25319 & 0.25319 & 0.25300 & 0.25299 & 0.25299 & 0.25056 & 0.25056 & 0.25056 & 0.24720 & 0.24728 & 0.24724 & 0.24338 & 0.24337 & 0.24337 \\
Bond(C1-H2) & 0.24133 & 0.24061 & 0.24060 & 0.24371 & 0.24367 & 0.24367 & 0.24752 & 0.24751 & 0.24751 & 0.24644 & 0.24653 & 0.24649 & 0.24338 & 0.24353 & 0.24353 \\
Bond(H3-H4) & 0.14036 & 0.14036 & 0.14036 & 0.14662 & 0.14662 & 0.14662 & 0.17622 & 0.17622 & 0.17623 & 0.20573 & 0.19909 & 0.19912 & 0.23092 & 0.23072 & 0.23074 \\
Bond(C1-H3) & 0.13248 & 0.13252 & 0.13252 & 0.12100 & 0.12100 & 0.12100 & 0.07971 & 0.07971 & 0.07971 & 0.04911 & 0.05217 & 0.05207 & 0.02504 & 0.02509 & 0.02508 \\
\hline
\multicolumn{16}{c}{Laplacian} \\ \hline  
nucleus (C) & -3.39E+06 & -3.39E+06 & -3.39E+06 & -4.09E+06 & -4.09E+06 & -4.09E+06 & -3.97E+06 & -3.97E+06 & -3.97E+06 & -3.19E+06 & -3.19E+06 & -3.19E+06 & -4.28E+06 & -4.28E+06 & -4.28E+06 \\
nucleus (H) & 2.19E+01 & 2.17E+01 & 2.17E+01 & 2.04E+01 & 2.04E+01 & 2.04E+01 & 1.77E+01 & 1.77E+01 & 1.76E+01 & 1.86E+01 & 1.86E+01 & 1.86E+01 & 1.41E+01 & 1.41E+01 & 1.41E+01 \\
nucleus (H) & 1.98E+01 & 1.99E+01 & 1.99E+01 & 1.24E+01 & 1.24E+01 & 1.24E+01 & 3.10E+01 & 3.10E+01 & 3.10E+01 & 1.53E+01 & 1.48E+01 & 1.48E+01 & 1.18E+01 & 1.18E+01 & 1.18E+01 \\
nucleus (H) & 1.32E+01 & 1.33E+01 & 1.33E+01 & 1.74E+01 & 1.74E+01 & 1.74E+01 & 1.57E+01 & 1.57E+01 & 1.57E+01 & 1.29E+01 & 1.25E+01 & 1.25E+01 & 1.42E+01 & 1.42E+01 & 1.42E+01 \\
nucleus (H) & 1.59E+01 & 1.59E+01 & 1.59E+01 & 2.49E+01 & 2.49E+01 & 2.49E+01 & 2.59E+01 & 2.59E+01 & 2.59E+01 & 1.28E+01 & 1.28E+01 & 1.28E+01 & 4.26E+01 & 4.25E+01 & 4.25E+01 \\
nucleus (H) & 1.59E+01 & 1.59E+01 & 1.59E+01 & 2.49E+01 & 2.49E+01 & 2.49E+01 & 2.59E+01 & 2.59E+01 & 2.59E+01 & 1.28E+01 & 1.28E+01 & 1.28E+01 & 4.25E+01 & 4.25E+01 & 4.25E+01 \\
Bond(C1-H5) & -7.42E-01 & -7.43E-01 & -7.43E-01 & -7.55E-01 & -7.55E-01& -7.55E-01 & -7.59E-01 & -7.59E-01 & -7.59E-01 & -7.65E-01 & -7.75E-01 & -7.78E-01 & -7.51E-01 & -7.02E-01 & -7.02E-01 \\
Bond(C1-H6) & -7.42E-01 & -7.44E-01 & -7.43E-01 & -7.55E-01 & -7.55E-01& -7.55E-01 & -7.59E-01 & -7.59E-01 & -7.59E-01 & -7.65E-01 & -7.76E-01 & -7.77E-01 & -7.04E-01 & -7.06E-01 & -7.06E-01 \\
Bond(C1-H2) & -6.72E-01 & -6.74E-01 & -6.73E-01 & -6.98E-01 & -6.99E-01& -6.98E-01 & -7.41E-01 & -7.42E-01 & -7.43E-01 & -7.59E-01 & -7.57E-01 & -7.59E-01 & -7.00E-01 & -7.51E-01 & -7.51E-01 \\
Bond(H3-H4) & -7.98E-02 & -7.63E-02 & -7.64E-02 & -1.09E-01 & -1.08E-01& -1.08E-01 & -3.07E-01 & -3.06E-01 & -3.07E-01 & -4.70E-01 & -4.50E-01 & -4.51E-01 & -6.73E-01 & -6.73E-01 & -6.73E-01 \\
Bond(C1-H3) & -1.06E-01 & -1.08E-01 & -1.08E-01 & -8.09E-02 & -7.99E-02& -7.99E-02 & 4.17E-02 & 4.17E-02 & 4.17E-02 & 9.70E-02 & 8.02E-02 & 8.04E-02 & 7.85E-02 & 7.85E-02 & 7.83E-02 \\
\hline
\end{tabular}}
\phantomsection
\label{tab:density-SI}
\end{table}

\begin{table}[H]
\centering
\scriptsize
\setlength{\tabcolsep}{5pt}
\caption{Mulliken population analysis of the CH$_5^+$ dissociation geometries with VQE and VQE* compared to CISD. The k-UpCCGSD ansatz, active space (4,4)  and STO-3G basis were used.}
\renewcommand{\arraystretch}{1.1}
\scalebox{0.82}{%
\begin{tabular}{c|ccc|ccc|ccc|ccc|ccc}
\hline
\textbf{Átomo} & \multicolumn{3}{c}{\textbf{1.3}} & \multicolumn{3}{c}{\textbf{1.4}} & \multicolumn{3}{c}{\textbf{1.6}} & \multicolumn{3}{c}{\textbf{1.8}} & \multicolumn{3}{c}{\textbf{2.1}}  \\ \hline
 & VQE & VQE* & CISD & VQE & VQE* & CISD & VQE & VQE* & CISD & VQE & VQE* & CISD & VQE & VQE* & CISD \\
\hline
\multicolumn{16}{c}{\textbf{Mulliken Charges}} \\ \hline
C1   & -0.19441 & -0.19697 & -0.19699 & -0.15009 & -0.15027 & -0.15028 & -0.00506 & -0.00512 & -0.00514 & 0.10945 & 0.06067 & 0.05856 & 0.16188 & 0.16025 & 0.16036 \\
H2  & 0.22923  & 0.23502  & 0.23505  & 0.22838  & 0.22873  & 0.22878  & 0.22398  & 0.22405  & 0.22409  & 0.23040 & 0.22961 & 0.23048 & 0.24238 & 0.24246 & 0.24248 \\
H3  & 0.26504  & 0.26167  & 0.26163  & 0.24330  & 0.24310  & 0.24300  & 0.16841  & 0.16837  & 0.16833  & 0.09828 & 0.12376 & 0.12375 & 0.05341 & 0.05410 & 0.05401 \\
H4  & 0.26865  & 0.26597  & 0.26598  & 0.24535  & 0.24519  & 0.24525  & 0.16896  & 0.16894  & 0.16896  & 0.09855 & 0.12404 & 0.12368 & 0.05372 & 0.05440 & 0.05431 \\
H5  & 0.21575  & 0.21715  & 0.21716  & 0.21654  & 0.21663  & 0.21662  & 0.22186  & 0.22188  & 0.22188  & 0.23166 & 0.23096 & 0.23177 & 0.24431 & 0.24440 & 0.24442 \\
H6  & 0.21575  & 0.21715  & 0.21716  & 0.21654  & 0.21663  & 0.21662  & 0.22186  & 0.22188  & 0.22188  & 0.23166 & 0.23096 & 0.23177 & 0.24431 & 0.24440 & 0.24442 \\
\hline
\multicolumn{16}{c}{\textbf{Mulliken Populations}} \\ \hline
C (1s)   & 1.9928  & 1.9928  & 1.9928  & 1.99302 & 1.99302 & 1.99302 & 1.99369 & 1.99369 & 1.99369 & 1.99413 & 1.99410 & 1.99411 & 1.99442 & 1.99442 & 1.99442 \\
C (2s)   & 1.30083 & 1.30083 & 1.30083 & 1.31202 & 1.31202 & 1.31201 & 1.34834 & 1.34835 & 1.34835 & 1.37505 & 1.37173 & 1.37297 & 1.39436 & 1.39449 & 1.39454 \\
C (2px)  & 1.12476 & 1.12725 & 1.12726 & 1.13469 & 1.13486 & 1.13485 & 1.14889 & 1.14890 & 1.14891 & 1.15820 & 1.15819 & 1.15819 & 1.16995 & 1.16995 & 1.16995 \\
C (2py)  & 0.61450 & 0.61458 & 0.61459 & 0.55053 & 0.55055 & 0.55058 & 0.35649 & 0.35656 & 0.35656 & 0.20322 & 0.25536 & 0.25620 & 0.11067 & 0.11219 & 0.11202 \\
C (2pz)  & 1.16152 & 1.16151 & 1.16151 & 1.15983 & 1.15982 & 1.15982 & 1.15765 & 1.15762 & 1.15762 & 1.15995 & 1.15995 & 1.15997 & 1.16872 & 1.16870 & 1.16871 \\
H2 (1s)  & 0.77077 & 0.76498 & 0.76495 & 0.77162 & 0.77127 & 0.77122 & 0.77602 & 0.77595 & 0.77591 & 0.76960 & 0.77039 & 0.76952 & 0.75762 & 0.75754 & 0.75752 \\
H3 (1s)  & 0.73496 & 0.73833 & 0.73837 & 0.75670 & 0.75690 & 0.75700 & 0.83159 & 0.83163 & 0.83167 & 0.90172 & 0.87624 & 0.87625 & 0.94659 & 0.94590 & 0.94599 \\
H4 (1s)  & 0.73135 & 0.73403 & 0.73402 & 0.75465 & 0.75481 & 0.75475 & 0.83104 & 0.83106 & 0.83104 & 0.90145 & 0.87596 & 0.87632 & 0.94628 & 0.94560 & 0.94569 \\
H5 (1s)  & 0.78425 & 0.78285 & 0.78284 & 0.78346 & 0.78337 & 0.78338 & 0.77814 & 0.77812 & 0.77812 & 0.76834 & 0.76904 & 0.76823 & 0.75569 & 0.75560 & 0.75558 \\
H6 (1s)  & 0.78425 & 0.78285 & 0.78284 & 0.78346 & 0.78337 & 0.78338 & 0.77814 & 0.77812 & 0.77812 & 0.76834 & 0.76904 & 0.76823 & 0.75569 & 0.75560 & 0.75558 \\
\hline
\end{tabular}
}
\phantomsection
\label{tab:mulliken-SI}
\end{table}

\begin{figure}[H]
    \centering
    \includegraphics[width=1.0\textwidth]{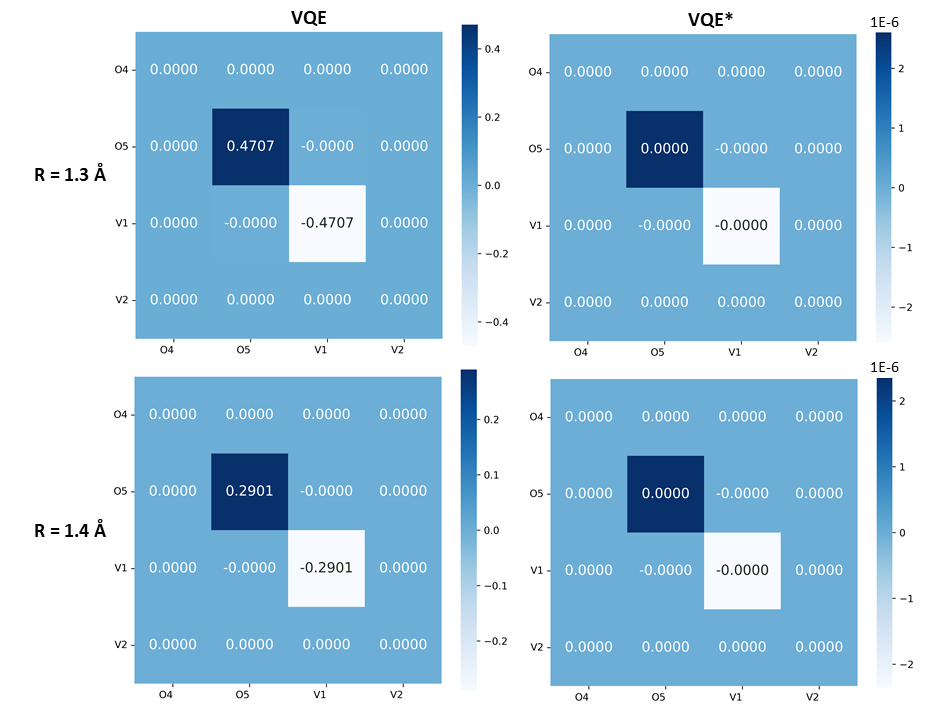}
    \caption{Difference in 1-RDM obtained with VQE and VQE* compared to CISD. The GateFabric \textit{ansatz}, the active space (2,2) and the STO-3G basis were used. O4 and O5 refer to the last two occupied orbitals, while V1 and V2 are the first two virtual orbitals.Here, O5 and V1 are the active orbitals.}
    \phantomsection
    \label{fig:rdm1-dif-gf}
\end{figure}

\end{document}